\documentclass[journal,12pt,onecolumn,letterpaper]{IEEEtran}
\usepackage{arxiv}
\usepackage[utf8]{inputenc} 
\usepackage[T1]{fontenc}    
\usepackage{hyperref}       
\usepackage{url}            
\usepackage{booktabs}       
\usepackage{amsfonts}       
\usepackage{nicefrac}       
\usepackage{microtype}      
\usepackage{lipsum}
\usepackage{amsmath}
\usepackage{amssymb}
\usepackage{lscape}
\usepackage{color}
\usepackage{array}
\usepackage{adjustbox}
\usepackage{tabularray}
\usepackage{graphicx}
\usepackage{subfigure}
\usepackage{rotating}
\usepackage{siunitx}
\usepackage{booktabs}

\title{FlexiChain 2.0: NodeChain Assisting Integrated Decentralized Vault for Effective Data Authentication and Device Integrity in Complex Cyber-Physical Systems}

\author{
	Ahmad J. Alkhodair\\
	Department of Computer Science and Engineering,\\
University of North Texas, \\
TX 76207, USA. \\
	\texttt{AhmadAlkhodair@my.unt.edu} \\
	\And
	Saraju P. Mohanty\\
	Department of Computer Science and Engineering,\\
	University of North Texas. \\
	TX 76207, USA. \\
	\texttt{saraju.mohanty@unt.edu}\\
		\And
Elias Kougianos \\
	Department of Electrical Engineering,\\
University of North Texas. \\
TX 76207, USA. \\
	\texttt{elias.kougianos@unt.edu} \\
}

\begin{document}

\maketitle

\begin{abstract}
Distributed Ledger Technology (DLT) has been introduced using the most common consensus algorithm either for an electronic cash system or a decentralized programmable assets platform which provides general services. Most established reliable networks are unsuitable for all applications such as smart cities applications, and, in particular, Internet of Things (IoT) and Cyber Physical Systems (CPS) applications. The purpose of this paper is to provide a suitable DLT for IoT and CPS that could satisfy their requirements.  
The proposed work has been designed based on the requirements of Cyber Physical Systems. FlexiChain is proposed as a layer zero network that could be formed from independent blockchains. Also, NodeChain has been introduced to be a distributed (Unique ID) UID aggregation vault to secure all nodes' UIDs. Moreover, NodeChain is proposed to serve mainly FlexiChain for all node security requirements. NodeChain targets the security and integrity of each node. Also, the linked UIDs create a chain of narration that keeps track not merely for assets but also for who authenticated the assets.  
The security results present a higher resistance against four types of attacks. Furthermore, the strength of the network is presented from the early stages compared to blockchain and central authority. 
FlexiChain technology has been introduced to be a layer zero network for all CPS decentralized applications taking into accounts their requirements. FlexiChain relies on lightweight processing mechanisms and creates other methods to increase security.
\end{abstract}


\section{Introduction}
\label{sec:Introduction}

A person's digital identity is something they utilize everywhere they go online, including at home, in the office, and when using any number of other services and devices. Everything that we say, do, or experience—-from buying concert tickets to checking into a hotel to placing a lunch order-—makes up our lives. In the present, our digital personae and interactions are owned and managed by third parties, some of whom we may not even know about.

DLT, or distributed ledger technology, is a set of protocols and systems that facilitate immutable record keeping across a network that spans many organizations or physical locations, and permits simultaneous record access, validation, and updates. Blockchain Technology (BT) is the first Distributed Ledger System (DLS) to solve the problem of Double Spending (DS) attack with the cryptocurrency market and was initially introduced as the technical background of the Bitcoin system \cite{PoW_Bitcoin}. One of the essentials of blockchain is that, because the distributed ledger is immutable, it makes it possible for parties to a exchange and others to act as stakeholders to build trust among untrusted entities in a decentralized approach. Even though the beginning of blockchain was to support cryptocurrencies, it has became a technology for distributed systems that has inspired and driven a shift from centralized to decentralized and dynamic system architectures \cite{Everything_Blockchain}. The blockchain-based architecture is decentralized and open because it is run by a number of distributed nodes. Each of these nodes has a copy of the cryptographically chained grouped data (Blocks), which are organized upon time. These blocks are agreed upon by the blockchain nodes using consensus protocols.

Data tracing, data storage, data protection, and data decentralization are only few of the goals that have been investigated and solved with the help of the blockchain. Since then experts, researchers, and businesses have begun implementing blockchain technology in various sectors, including healthcare, transportation, agriculture, and government. There may be positive and negative effects on performance from implementing a blockchain in a CPS or IoT setting. For instance, the technique necessitates devices with powerful capabilities since it relies on high-volume processing and complex calculations.

In a decentralized ecosystem, each node is represented by three values: a public key used to represent the user in the network, a private key that should be kept secret and not shared with anyone because it is the value that gives the user the ability to claim transactions and sign them, and the address a shortened value of the public key that represents that value. The generation of these keys is automated in the network, however they are requested and kept manually.

This approach of generating an identity is adequate if all ledgers are operating over one blockchain as a unified layer. Despite the fact that there are many others which provides valuable services, to join them we have to create other credentials. The idea of keeping multiple digital signatures for various network is impractical. This issue is raised due to the absence of a layer zero in the blockchain market that could link all blockchains together. The problem extends to include other problems if resource constrained nodes are part of the network. Primary among these are decentralized IoT and CPS applications.  
Some security levels exist in some high resource nodes. Such a security enclave that could create a UID for authentication is incompatible to be used in authentication processes for decentralized networks. Also, some resource constrained devices do not have that hardware as part of their circuitry to keep their own Unique Identification (UID).

If DLT is the main framework for CPS and IoT applications, the nodes within that network will acquire the same credentials to engage in the decentralized work. However, the limited capabilities nodes will not be able to keep their own credentials. That is the main reason why all proposals in this field rely on private networks in which all participants are predefined to the network. This raises the need to have a certain method that could provide credentials to resource constrained nodes without a burden on computation with ensuring their security and integrity.

Moreover, each IoT or CPS node might need to have more than just one credential to operate over the network which might complicate the notion of applying DLT based applications. The integrity and security of nodes are an essential part since using well known consensus algorithms is inefficient in resource constrained nodes because it will consume all its resources and might create  new risks to the network. 

However, security hardware could resolve the authentication process and create a unified physical layer0 unique identification to be used in all distributed ledger networks authentication of which will reduce the computation which will result in decreasing the cost and increase sustainability and scalability. Immutability of any distributed ledger technology relies on the huge amount of computation or process power wasted to mine or validate a block or transaction. Immutability is an important factor in Blockchain technology and any other distributed ledger from which the network increase its reliability and transparency. Decreasing these massive operations should be compensated with other security hardware layers to keep the network secure and immutable and CPS, IoT compatible.

\begin{figure}[htpb]
	\centering
	\includegraphics[width=0.90\textwidth]{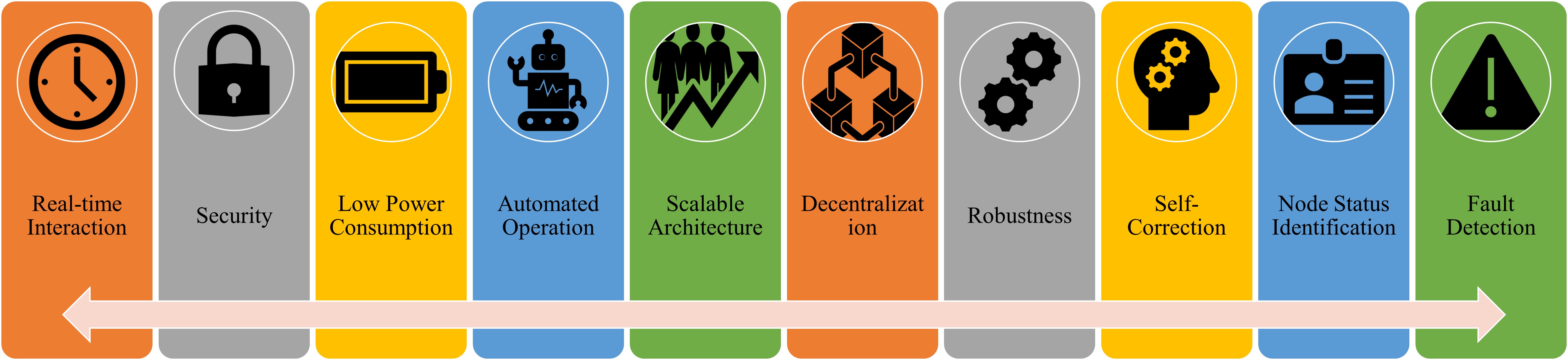}
	\caption{Examples of Cyber Physical System and Internet of Things Distributed Ledger Technology Based Requirements}
	\label{FIG:CPS_Requirements_MultiBlockchain}
\end{figure}

Each area of decentralized applications needs a certain design of a distributed ledger that could satisfy its requirements, as shown in figure \ref{FIG:CPS_Requirements_MultiBlockchain}. Otherwise, using blockchain as a platform might cause an issue rather than providing a solution. 
Observing the challenges above, this paper proposes a solution that provides a layer zero network that unifies the access of all networks in one credential while increasing the privacy and security of exchanged assets and nodes of which their integrity is supervised by our novel proposed method.       

The organization of this paper is as follows: Section 2 \ref{sec:NovelContributions} summarizes the novel contributions of this paper. Section 3 \ref{sec:BackgroundPriorRelatedResearchWorks} presents  summarizes related research and provides comparative perspectives of our work compared to published related research. Section 4 \ref{sec:ProposedMethod} discusses the proposed FlexiChain and its consensus algorithm. Section 5 \ref{sec:Implementation} presents experimental results. Sections 6 and 7 \ref{sec:Conclusions} conclude the paper and present directions for future research.

\section{Novel Contributions of the Current Paper} 
\label{sec:NovelContributions}

This section covers contributions of this paper and highlights the novel proposed work. This paper addresses several problems in DLT based complex applications such as CPS and the IoT. Multiple identities for each node are complicated since all keys should be kept privately and not exposed to the public. Moreover, the approach of creating and securing them is incompatible with Resource Constrained Nodes (RCN). In addition to utilizing power-consuming protocols to reach a consensus over a block and authenticate it, which is also time consuming due to scalability problems. Voting and puzzle related consensus algorithms have their advantages. However, in some applications theses advantages could create other challenges. For instance, Proof of Work (PoW) is an effective mechanism in ledger immutability; however, it is time and power consuming and requires high resources. Proof of Stake (PoS) is an effective protocol in terms of low latency; however, a fifty one percent attack is more probable than PoW. Also, it requires high resource nodes. Some other DLT protocols are efficient as well, such as Hashgraph technology. However, the redundancy is there since they rely on voting and a probabilistic mechanism to reach a consensus.  

The importance of digital identifiers for physical objects in the next generation of CPS suggests our previous work NodeChain and FlexiChain \cite{Alkhodair2021}, could significantly broaden their potential use cases, bringing them into fields like smart agriculture, smart supply chain management, and smart transportation. NodeChain, which provides a novel ledger that includes the virtual existence of physical entities in an independent ledger that could serve as the assisted off-chain blockchain for FlexiChain and others. Adding the NodeChain \cite{Alkhodair2021} to MultiChain \cite{Alkhodair2020} further speeds up the authentication procedure and adds an extra layer of security to the entry point for every device connecting to the network.

\begin{figure*}[htpb]
	\centering
	\includegraphics[width=0.90\textwidth,trim={0cm 0cm 0cm 0cm}]{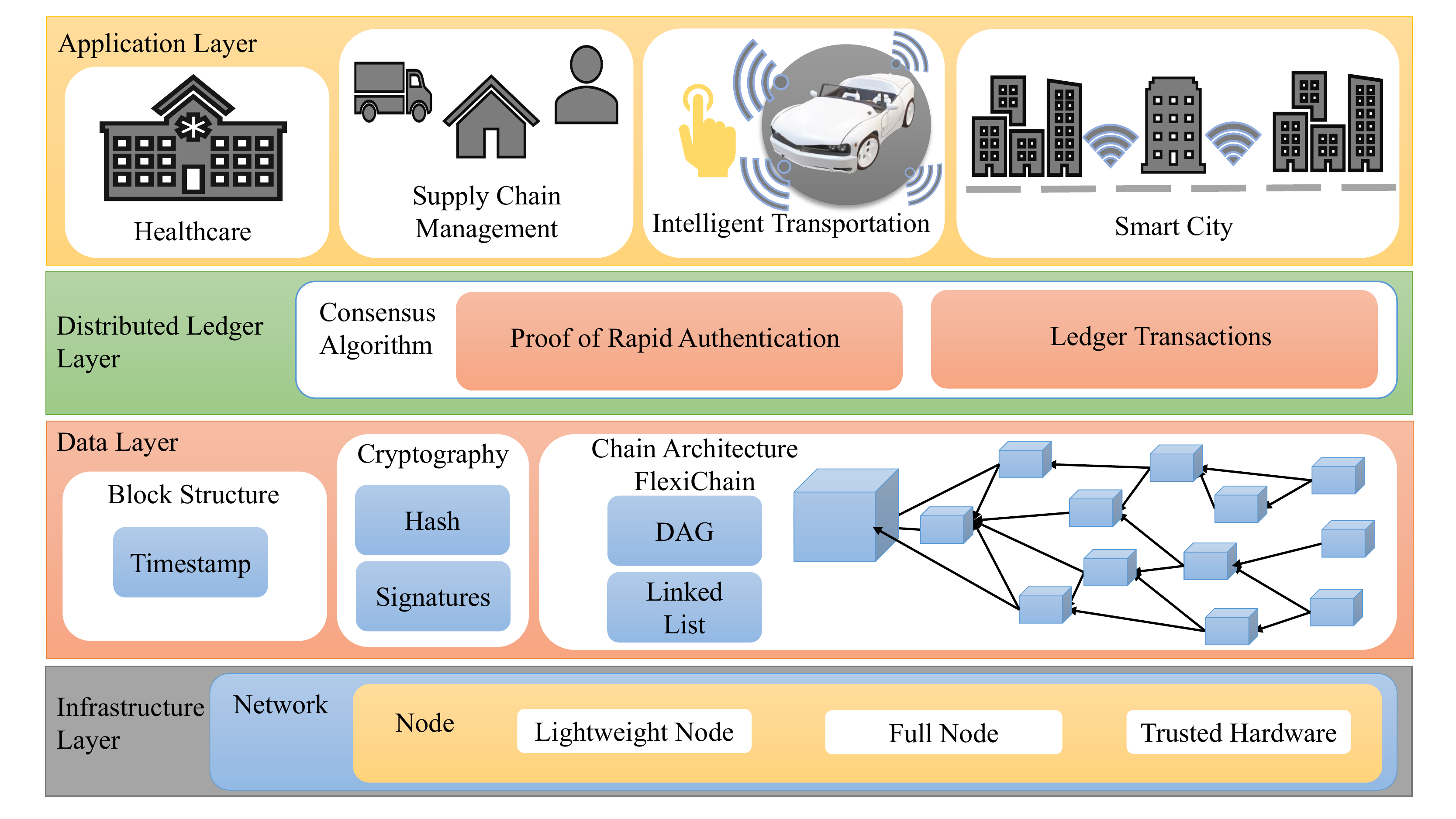}
	\caption{Levels of FlexiChain} 
	\label{FIG:Introduction_Flexichain2_FrameworkMultiBlockchain} 
\end{figure*}

The NodeChain is a distributed ledger that monitors and authenticates all the virtual copies of the physical nodes in a network using virtual copy headers and Unique Identifications (UIDs). NodeChain is a blockchain-based network method which utilizes the original blockchain structure. The objective of this off-chain assisted ledger is to create and link digital UIDs of devices in order to make impersonating or attacking them very costly, as depicted in figure \ref{FIG:Introduction_Flexichain2_FrameworkMultiBlockchain}. This process covers node enrollment as well as their integrity and security in the network. 

This security method has been introduced particularly to direct communication and authentication, such as Machine to Machine (M2M), Vehicle to Vehicle (V2V), Vehicle to Machine (V2M), Human to Vehicle (H2V), and Human to Machine (H2M) for resource constrained nodes that operate over a decentralized network and exchange data on a public ledger (FlexiChain). NodeChain represents an assisting security layer that compensates the reduction in calculations to support and speed up the validation process in the distributed ledger instead of just relying just on the hashcash process or the probabilistic consensus algorithms that could burden the network due to their redundancy.

NodeChain is proposed to be used either as an assisted ledger to provide security to resource constrained nodes or as an authentication process for layer0 FlexiChain. The authentication process for digital assets in the network will be through the UID assigned for each node and derived from the pre-manufacturer enrollment process. The enrollment and authentication processes will be further explained in the proposed work section. NodeChain aims to provide hardware based security for all node types and capabilities in the network, and to unify all layer1 blockchains under one unified representation for all layer1 blockchains and to perform the validation and authentication process in authority without the need to have high resources, high stake or a probabilistic approach. Also, FlexiChain is a layer0 distributed ledger that combined the strength in security and immutability of the traditional blockchain (Bitcoin like) and the flexibility and scalability that been presented in Directed Cyclic Graph (DAG) approaches such as IOTA. This new novel ledger has been introduced to provide a layer0 ledger that could combine multiple DAG based blockchains and keep them strongly connected.

In figure \ref{FIG:Introduction_Flexichain2_FrameworkMultiBlockchain} the first level of the network is the nodes and their sensor and actuator in the scenario of an intelligent transportation system. The second level is the independent ledger that holds UIDs and parameters which are used to generate a UID, and how this independent ledger could be an assisted ledger. The third level presents how the Independent ledger is integrated with FlexiChain technology.


\begin{figure*}[htpb]
	\centering
	\includegraphics[width=0.90\textwidth,trim={0cm 0cm 0cm 0cm}]{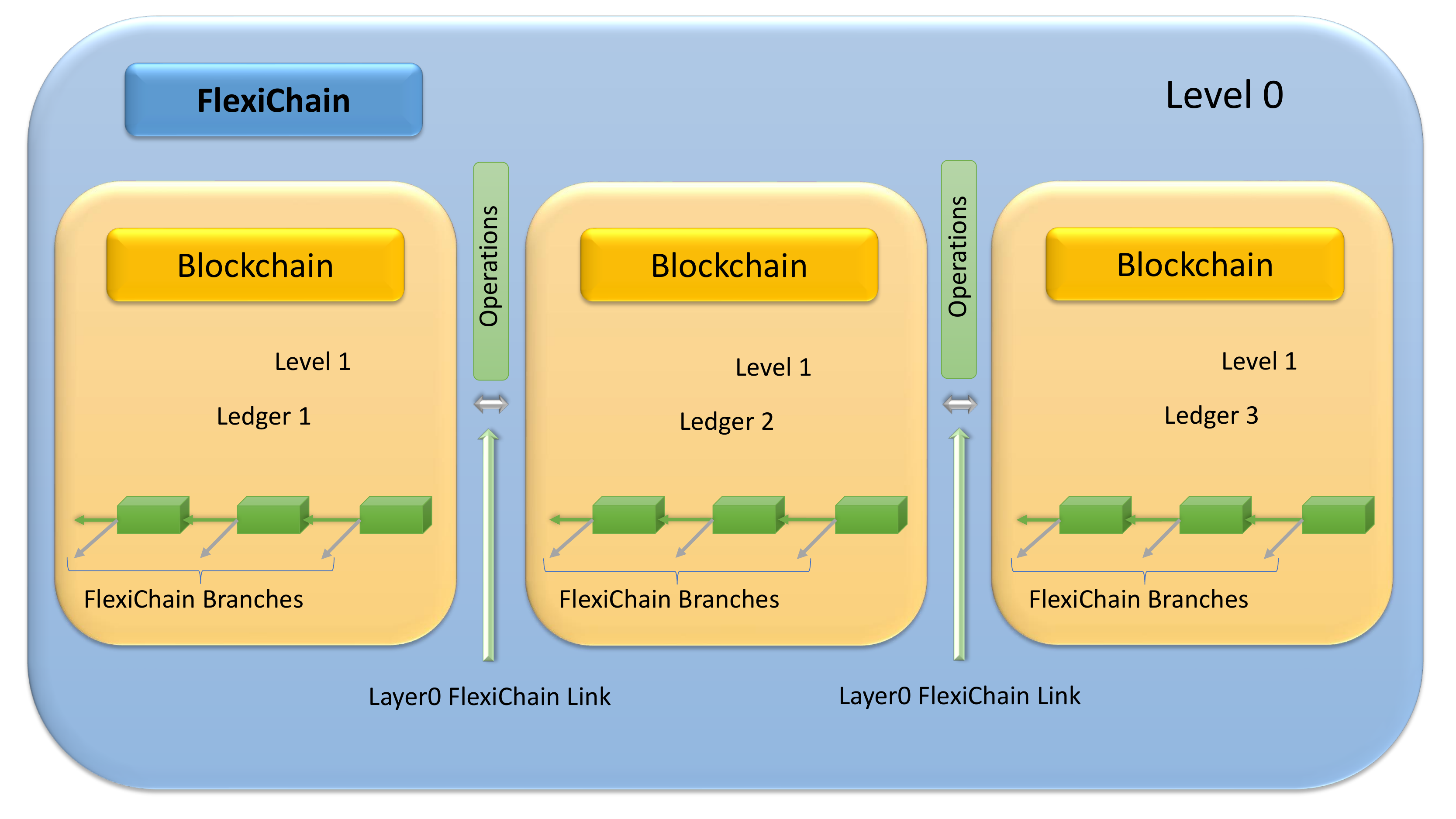}
	\caption{NodeChain Integrated With FlexiChain} 
	\label{FIG:Introduction_NodeChainInFlexiChain_layers} 
\end{figure*}

Our proposal targets all challenges mentioned above and introduces a network hardware unified enrollment for all branches in FlexiChain or provides an independent decentralized assisted credentials in an offline vault to other ledgers for security and authentication, as shown in figure \ref{FIG:Introduction_NodeChainInFlexiChain_layers}. The proposed novel contributions are summarized as follow:

\begin{enumerate}
	\item To the best of the authors' knowledge, the method is the first to use intrinsic and extrinsic nodes manufacturing parameters to create a unified unique identity for each node.
	\item To the best of our knowledge, this approach is the first to use trusted hardware module to perform initial network enrollment. 
	\item The proposed method creates a chain of narration once an enrolled node authenticates a block. A chain of narration is a novel delegation process to allow nodes to chain their UIDs and the authentication of the recent nodes is equivalent in all chains.  
	\item The network uses New Node State (NNS) to ensure that all nodes have updated their NodeChain.   
	\item NodeChain is used to monitor hardware changes in the network by observing each node virtual header. 
	\item NodeChain could be an offline support decentralized blockchain vault used for authentication for other ledgers. 
	\item It provides a unified ledger for aggregating node credentials instead of having multiple keys.   
\end{enumerate}

\section{Background and  Related Research}
\label{sec:BackgroundPriorRelatedResearchWorks}

Blockchain technology applications has been demonstrated in the Ethereum and Bitcoin cases \cite{Blockchain_framework} but they are not open to other use cases by design \cite{POS,Everything_Blockchain} since the technology lacks scalability in practically every aspect \cite{PoW_Bitcoin}. Tangle technology has proven to be the most effective as a blockchain replacement for Internet of Things (IoT) and Cyber Physical System (CPS) applications \cite{IOTAIOT}, and employs Directed Acyclic Graphs (DAG) and single transaction as the data architecture \cite{DAG1} \cite{DAG2} \cite{DAG3} \cite{DAG4}. Hashgraph technology that was deployed by Hedera has became an alternative due to its gossip protocol and DAG \cite{HashGraph}. Hedera is the only real-life  deployment example of the hashGraph technology protocol \cite{Hedera}. A DAG  architecture is used by Black-Lattice Technology \cite{LeMahieu2015}. In \cite{IoTex}, a customized version of the blockchain (consortium type) is presented to fit IoT case uses. A combination of Delegated Proof of Stake (DPoS), Practical Byzantine Fault Tolerance (pBFT), and Verifiable Random functions is used to produce Roll-DPoS that could fit IoT use-cases \cite{IoTex} \cite{VRF}. Similarly, The Nodle Network aims to provide IoT devices with a worldwide, simple-to-use, decentralized communication network \cite{Nodle}. The Internet of Services (IoS) \cite{IOST} is a fast network that deploys the Proof of Believability (PoB) which uses a believability score for selecting a validator. In \cite{PoActivityAndNovelConsensusBased} nodes that are not involved in the process of validating and broadcasting blocks will not be able to claim their share of the incentives pool, but those that are will be rewarded with a stake if they participate in the validation and broadcasting of blocks.
Table \ref{TBL:Established_Related_Blockchain_Vs_Tangle_Vs_Proposed} lists and compare some of the established related works to the current paper. 

\begin{table*}[htbp]
	\caption{A Comparative Perspective of Related Works Versus FlexiChain.}
	\label{TBL:Established_Related_Blockchain_Vs_Tangle_Vs_Proposed}
	\centering
	\scriptsize
	\begin{tabular}{|p{1.2cm}|p{1.2cm}|p{1.3cm}|p{1.2cm}|p{1.2cm}|p{1.5cm}|p{1.2cm}|p{1.3cm}|p{1.5cm}|p{1.2cm}|}
		\hline
		\textbf{Features}  & \textbf{Linked Lists} & \textbf{Registration} & \textbf{Type of Validation} & \textbf{Types of Nodes} & \textbf{Cryptography} & \textbf{Consensus} &\textbf{Power Consumption} & \textbf{Design Purpose} &\textbf{Block type}\\ 
		\hline
		\hline
		Blockchain (Bitcoin) \cite{Everything_Blockchain,PoW_Bitcoin} & Linked list of blocks  &  Manual  &  Mining & Traders and Miners & Digital Signatures & (PoW) & High & Electronic Cash System & One \\		
		\hline
		Tangle (IOTA) \cite{IOTAIOT,Tangle} & (DAG) of transactions & Manual & Mining & Traders and Coordinators & Digital Signatures & (PoW) & High & IoT micro payments & One\\
		\hline 
		HashGraph (Hedera) \cite{HashGraph,Hedera} & (DAG) of transactions hash & Manual & Virtual voting (witness) & Users & Digital Signatures & (ABFT) & Medium & Decentralized Applications & One \\
		\hline
		Blockchain (IoTx) \cite{IoTex}  & linked List of blocks & Manual & Validating & Users and Validators & Digital Signatures & (RDPoS) & Medium & IoT Applications & One \\
		\hline
		Blockchain (Nodle) \cite{Nodle} & linked list of Blocks & Manual & Validating & Users, Contributors, and Validators & Digital Signatures & (PoConnectivity) & Low & IoT Applications & One\\
		\hline 
		Blockchain (IOST) \cite{IOST} & linked list of blocks & Manual & Validating & Users and Validators & Digital Signatures & (PoB) & Medium & IoT services & One\\
		\hline
		FlexiChain 1.0 \cite{Alkhodair2022}& DAG of Blocks & Trusted Hardware & Authentication & Users & All Nodes & Constructed and Trusted keys& (PoRa) & Low & Equal to Initiation\\
		\hline 
		FlexiChain 2.0 \textbf{Current Paper} & Layer 0 Ledger & Trusted Hardware & Authentication & Backup, Edge & All Nodes & Constructed and Trusted Keys & (NPoRa) & Low & \#of blockchains \\
		\hline 
	\end{tabular}
\end{table*} 

\subsection{Secure Enclave}
\label{subsec:SEP}

Manufacturers recently use security processors and utilize Systems On Chip (SoCs) that have a specialized security component called the Secure Enclave (SoCs) \cite{SEP}. The Secure Enclave isolates the main processor to protect sensitive user data even if the Application Processor core is corrupted. It has a boot ROM, an AES engine, and protected memory like the SoC. Although the Secure Enclave doesn't have storage, it can securely store information on associated storage separate from the Central Processing unit and operating on the system's NAND flash storage. This secure enclave comprises of a random number generator, AES engine, processor, PKA, and a memory. This kind of system provides a secure layer and supports the device from losing its sensitive data. The Secure Enclave Processor (SEP) is where fingerprint data, cryptography keys, and other sensitive identity representations reside and are isolated, which eliminates the CPU from accessing these data \cite{SecurityServiceEnginestoAccelerateEnclavePerformanceinSecureMulticoreProcessors}. The SEP is utilized for different services such as touch ID and has its own kernel, drivers, services and applications.

\subsection{Offline (Cold) Wallets}
\label{subsec:OCW}

Offline cold wallets serve as a storage to digital assets. This storage is an offline storage disconnected from the Internet. 
A software wallet can be used in offline mode, and is divided into two parts: one that stores the private keys and another that stores the public keys on the cloud. When a user makes a purchase using their online wallet, new, unsigned transactions are created and the user's address is sent to the buyer or seller. The offline wallet is where the transaction is taken to be signed using the private key. The final step is to send the signed transaction back to the digital wallet in order to have it broadcast to the network. Private keys are safe in an offline wallet since it is never connected to the Internet. An offline hardware wallet employs a dedicated device or smart card to store and secure one's private keys. One type of hardware wallet that makes use of a smart card to safeguard private keys is the Ledger USB Wallet.

\begin{table}[htpb]
	\caption{Comparative Perspective Between Security Hardware \& Proposed Work.}
\label{TBL:SecureEnclaveVersusOfflineCryptoWalletsVersussecurityHardwareBasedofflineledgerNodeChain}
	\centering
	\scriptsize
	\begin{tabular}{|p{3cm}| p{3.5cm}| p{3.5cm}| p{3.5cm}|}
		\hline
		\textbf{Features/Version}       	& \textbf{Secure Enclave} \cite{SecurityServiceEnginestoAccelerateEnclavePerformanceinSecureMulticoreProcessors}	& \textbf{Offline Crypto Wallets} \cite{Alkhodair2020} & \textbf{NodeChain} (Current)\\
		\hline
\hline		
		\textbf{Design Purpose}      	& \begin{enumerate}\item Hardware security layer \item Internal  System On Chip(SoC), Completely isolated \item Keeps Sensitive Data \end{enumerate} 
		& \begin{enumerate} \item Hardware security Crypto Storage \item USB \item keeps Tokens Secured \item external \& Offline \end{enumerate} 
		& \begin{enumerate} \item Hardware Security Layer Distributed Ledger Technology \item External Hardware \item Keeps Network Participants UIDs Secured \end{enumerate}\\
		\hline
		
		\textbf{Registration}           & Manufacturer Pre-designed and equipped  			  & Manufacturer designed and equipped  & Proposed as Manufacturer designed and equipped\\
		\hline
		
		\textbf{Authentication}         & \begin{enumerate}
			\item Digital Signatures
			\item Unique Identification (Randomly Generated) \end{enumerate}& Digital Signatures & \begin{enumerate}
			\item Digital Signatures
			\item Unique Identification Manufacturer Specification + PUF (Generated By SCRYPT)
		\end{enumerate} \\
		\hline
		
		\textbf{Type of Validation}     & Authentication                             & Authentication                 & Authentication and Lightweight Computation\\
		\hline

		\textbf{Resources Requirement}  & Used in High resources node as extra security layer			 & High \& Limited        & Use in resource Constrained node as extra security\\
		\hline
	\end{tabular}
\end{table}

In table.\ref{TBL:SecureEnclaveVersusOfflineCryptoWalletsVersussecurityHardwareBasedofflineledgerNodeChain}, the table clarifies the differences and similarities between the proposed Manufacturing Trusted Hardware used in the consensus algorithm and other examples for hardware security based concepts. The registration process for NodeChain will be through the manufacturers UID generated based on random unique number, manufacturing specification such as SRAM-PUF, and other nodes' UIDs fed to hash function to generate a new UID and others based on the requirements of the manufacturer. Tokenized UID will be in the NodeChain for faster authentication while the actual one will be in the secure hardware to keep it isolated.

\subsection{Prior Research in CPS}
\label{sec:Research_Related_Works}

The integration of DLTs to CPS and the IoT is very widespread among researchers and companies \cite{Blockchain_Applications}. This research trend takes place due to advantages and uniqueness of DLTs \cite{Blockchain_IoT_Edge}. The awareness of the necessity in customization based design purpose CPS and IoT requirements of DLTs is shown in figure \ref{FIG:CPS_Requirements_MultiBlockchain} \cite{Alkhodair2022}. 
In \cite{PUFchain} the proposed system is designed targeting security and speed. The paper proposes a novel consensus algorithm to reduce the power consumption burden over users and to speed up the whole validation process. In addition, it is using Physical Unclonable Functions (PUF) to ensure the integrity of nodes \cite{PuFChain2}. In \cite{Authenticity} consensus is achieved via Proof of Authenticity. In \cite {PoAh} \cite{PoAh2} a recent scalable consensus mechanism for device to device communication has been proposed. This is a lightweight protocol based on PoW but is faster and applicable to IoT environments since there are no nonce calculations. Another protocol built for business blockchain based on the IoT is targeting scalability and security \cite{PoBT}. The protocol consists of two stages: the trade verification, and the consensus formation. In \cite{POTrust}, ``Proof-of-Trust'' consensus (PoT) is a consensus protocol that uses RAFT leader election instead of traditional Paxos algorithm and Shamir's secret sharing algorithms to select transaction validators based on the trust values held by service participants.

Other related work for specific applications has been covered in tables \ref{TBL:AComparativePerspectiveofRelatedWorksinSmartHealthcare}, \ref{TBL:AComparativePerspectiveofRelatedWorksinSupplyChainManagement}, and \ref{TBL:AComparativePerspectiveofRelatedWorksinArtificialIntelligence}. to presents the need of customization to the technology based on the application desired.
\begin{table*}[htbp]
	\caption{A Comparative Perspective of Related Works in Smart Healthcare.}
	\label{TBL:AComparativePerspectiveofRelatedWorksinSmartHealthcare}
	\centering
	\scriptsize
	\begin{tabular}{|p{1.5cm}|p{2cm}|p{2cm}|p{1cm}|p{3.5cm}|p{3.5cm}|}
		\hline
		\textbf{Decentralized Applications (Dapps)}  & DLT Type & Consensus Algorithm & Resources Requirements & Modifications & Contributions\\
		\hline
		\hline
		\textbf{\cite{Fortified}} & IPFS Merkle Tree (Directed Acyclic Graph) & Cryptographic Signatures Authentication& Low & - IPFS used is a customized version of DLT. - No computation or calculations required. - Offline storage for actual blocks. - Online storage for only hashes. & Authors could use a customized DLT to overcome conventional DLT issues to suit their application. \\		
		\hline
		\textbf{\cite{GlobalChain}} & Blockchain (Linked List) & Not Mentioned & NA & Creating global ledger for medical records. Real-time not needed. Security, Privacy, interoperability targeted and provided & Creating a unified ledger for epidemic medical records
		Scalability issues encountered \\
		\hline 
		\textbf{\cite{DAAC}} & Blockchain (Linked List) & Not Specified, However, mentioned suitable choices such as Ethereum and Hyper Fabric Ledger& Based on choices: High & No modification to the core of the technology. However, a proposal of a migration system to a DLT with choices given  & Migration solution to speed up the adaptation of DLT \\
		\hline
		\textbf{\cite{EPOW_HealthRecords}} & Blockchain (Linked List) & Enhanced Proof of Work (PoW)& High & Customized version of PoW, that stores encrypted actual data in ledger & Increased privacy\\
		\hline
		\textbf{\cite{ABlockchain_empoweredFederatedLearninginHealthcare_basedCyberPhysicalSystems}} & Blockchain (Linked List) & Validation turns over participants to accumulate transactions based importance voting system  & Medium & Instead of relying on Stake or processing power, all full node participants get a turn to validate a block & create a reward system to encourage medical entities to honestly participate in a decentralized manner \\
		\hline 
		\textbf{\cite{smartHealthcareDiseaseManagment}} & Blockchain (Linked List) & Proof of Authority (PoA) & low & Using two types of ledgers: local (IPFS) and public (Blockchain ledger) & Framework that secure data and ensure privacy locally, and public ledger for interoperability\\
		\hline 		
	\end{tabular}
\end{table*}    

\begin{table*}[htbp]
	\caption{A Comparative Perspective of Related Works in Supply Chain Management (SCM)}
	\label{TBL:AComparativePerspectiveofRelatedWorksinSupplyChainManagement}
	\centering
	\scriptsize
	\begin{tabular}{|p{1.5cm}|p{2cm}|p{2cm}|p{1cm}|p{3.5cm}|p{3.5cm}|}
		\hline
		\textbf{Decentralized Applications (Dapps)}  & DLT Type & Consensus Algorithm & Resources Requirements & Modifications & Contributions\\
		\hline
		\hline
		\textbf{\cite{PharmaChain}} & Blockchain Technology (Linked List) & Proof of Authority (PoA) & Low - Medium  & Automation by smart contracts. And, Counterfeit detection. Real-time interactions   & Authors used smart contract to automate the operations. Also, Counterfeit detection. Tracking system. \\		
		\hline
		\textbf{\cite{CryptoPharmacy}} & Blockchain (Linked List) & Proof of Importance (PoI) & Low - Medium & Application's operations smart contract based. Using PoI modified version of PoS for better node evaluation. Operation over a public ledger & Mobile app. PoI instead of well known consensuses. \\
		\hline 
		\textbf{\cite{GaRuDa}} & IPFS Merkle Tree (Directed Acyclic Graph) & Cryptographic Signatures Authentication & Low- Medium & Authors could use a customized DLT to overcome conventional DLT issues to suit their application.& - IPFS used is a customized version of DLT. - No computation or calculations required. - Offline storage for actual blocks. - Online storage for only hashes.\\
		\hline
		\textbf{\cite{Blockchainseverywhereausecaseofblockchaininthepharmasupplychain}} & Blockchain (Linked List) & Proof of Work (PoW) & High & No changes in core or smart contracts use & Immutability that acquired from using PoW.\\
		\hline
		\textbf{\cite{TraceabilityofcounterfeitmedicinesupplychainthroughBlockchain}} & Blockchain (Linked List) & Cryptographic Signatures Authentication & Low - Medium & Instead of relying on Stake or processing power, all full node participants authenticates using digital signatures. & Privacy. Low computation \\
		\hline 
		\textbf{\cite{Drugledger}} & Blockchain (Linked List) & Proof of Work (PoW) & High & No core changes in the technology & Security Immutability \\
		\hline 		
	\end{tabular}
\end{table*}    

\begin{table*}[htbp]
	\caption{A Comparative Perspective of Related Works in Artificial Intelligence (AI)}
	\label{TBL:AComparativePerspectiveofRelatedWorksinArtificialIntelligence}
	\centering
	\scriptsize
	\begin{tabular}{|p{1.5cm}|p{2cm}|p{1.7cm}|p{1.8cm}|p{3.5cm}|p{3.5cm}|}
		\hline
		\textbf{Decentralized Applications (Dapps)}  & DLT Type & Consensus Algorithm & Resources Requirements & Modifications & Contributions\\
		\hline
		\hline
		\textbf{\cite{Blockchain_Based_DeepLearning_CPS}} & IPFS Merkle Tree (Directed Acyclic Graph) & Cryptographic Signatures Authentication& Low & - IPFS used is a customized version of DLT. - No computation or calculations required. - Offline storage for actual blocks. - Online storage for only hashes. & Authors could use a customized DLT to overcome conventional DLT issues to suit their application. \\		
		\hline
		\textbf{\cite{DeepReinforcement_Edge_Blockchain}} & Blockchain (Linked List) & Not Mentioned & NA & Creating global ledger for medical records. Real-time not needed. Security, Privacy, interoperability targeted and provided & Creating a unified ledger for epidemic medical records
		Scalability issues encountered \\
		\hline 		
	\end{tabular}
\end{table*}

\subsection{Related Works in Intelligent Transportation}

In this section,Intelligent Transportation Systems (ITS) will be discussed and related work will be covered since our proposed FlexiChain 2.0 is mainly targeting DLT based ITS. The Internet of Vehicles (IoV) is the interconnection among each Internet enabled vehicle to collect and analyze data, and provide  feedback \cite{TactileInternetforAutonomousVehiclesLatencyandReliabilityAnalysis} \cite{ArtificialIntelligenceEmpoweredEdgeComputingandCachingforInternetofVehicles}. The IoV has the potential to become the next major trend due to advancements in satellite communications, Artificial Intelligence (AI) and CPS \cite{JointPricingandPowerAllocationforMultibeamSatelliteSystemsWithDynamicGameModel} \cite{SpectralEfficiencyEnhancementinSatelliteMobileCommunicationsAGameTheoreticalApproach}. Vehicle intelligence and automation will increase in the coming decade. The growth in vehicles will produce a huge amount of data, based on the data accuracy level and traffic management. Latency, complexity and IoV requirements will face substantial challenges if the operations are over the conventional paradigm the IoT. Moreover, it is also challenging to ensure compatibility and interoperability between IoV components offered by different service providers. In order to support the IoV's expansion and enable ITS's full potential, its data exchange and storage infrastructure must be decentralized, distributed, inter-operable, flexible, and scalable \cite{BlockchainfortheInternetofVehiclesTowardsIntelligentTransportationSystemsASurvey}. In \cite{BlockchainforSecureandEfficientDataSharinginVehicularEdgeComputingandNetworks} consortium blockchain is used to set up a secure, distributed data management system based on blockchain within the vehicular edge computing networks. The proposed system would benefit in two ways from using smart contracts. In \cite{DrivMan} DrivMan, a blockchain-based solution for cars that helps with trust management, data provenance, and privacy through a smart contract, is presented. It uses physically unclonable functions (PUF), and public-key infrastructure (PKI). In \cite{TraceableandAuthenticatedKeyNegotiationsviaBlockchainforVehicularCommunications} the main proposal of this security scheme in the paper contributes a new way to negotiate keys that can be tracked and authenticated. In particular, this plan is meant to solve a number of problems, such as security, trust, and monitoring of shared data. In \cite{ASecureandEfficientBlockchainBasedDataTradingApproachforInternetofVehicles} the blockchain is used to solve the problems that come up when trading data, such as a lack of transparency and traceability and unauthorized changes to data. It works like a consortium, where a group of local aggregators work together to audit and verify transactions. In \cite{ComputingResourceTradingforEdgeCloudAssistedInternetofThings} ways were suggested to solve problems with resource trading in edge-cloud-based systems, such as making sure bids are honest and letting both buyers and sellers trade. In \cite{BlockchainEmpoweredResourceTradinginMobileEdgeComputingandNetworks} a D2D-ECN solution for trading resources and assigning tasks that use blockchain, smart contracts, edge computing, and device-to-device communication is proposed. In \cite{PrivacyPreservingSmartParkingSystemUsingBlockchainandPrivateInformationRetrieval} a blockchain-based system that lets car drivers search for and book parking spots ahead of time in a way that is decentralized and protects their privacy is given.

There are many applications related to ITS. Direct communication between things and machines is one of the most applicable uses to employ Distributed Ledger Technology (DLT). Digital assets or payment systems are other applications that could use the advantages of distributed ledger technology to secure asset exchanges among users. This is an inter-operational ledger that could serve multiple agents in a concurrent way and secure their data using the unique characteristics of DLT.
	
Related works have shown that customization and modification in the current DLT architecture is important to fit the purpose of using it since it does not fit all applications \cite{WhenDoWeNeedAblockchain}. For example, IoT applications require fast responses to keep the whole operations run smoothly which will not work if the bitcoin operation of the blockchain is running. Cyber Physical Systems require real time responses to allow real time interactions between nodes which is hard to achieve based on what already existed. It is obvious from the related works discussed in this section that the parts of any DLT, such as consensus algorithms, digital signatures, hash functions, types of users, and validation types are all important in the design step. The most important parts of an application that is Internet of Things (IoT) or Cyber Physical System (CPS) related, are the DLT type and consensus algorithm since they are what makes the whole ledger decentralized, operable, immutable, and motivate users to use it \cite{Dedeoglu2020,IoT_meets_Blockchain}. 

\section{The Proposed FlexiChain 2.0}
\label{sec:ProposedMethod}

This section of the proposed framework highlights the enrollment that construct the base of this technology  and authentication processes that rely on the base. Also, the consensus mechanism is discussed. In FlexiChain 2.0, NodeChain is proposed as assisted distributed vault that can be used by other networks as a Digital Identity aggregator. Also, it is proposed as an integrated blockchain for FlexiChain authentication. 
In this section, an explanation is given of FlexiChain and how the NodeChain is integrated. Also, how NodeChain could serve as a support authentication ledger.

\subsection{Proposed Architecture}
\label{subsec:OverallArchitectureDiscussion}

\subsubsection{Ledger} \label{subsec:Ledger}

The FlexiChain ledger is based on multiple accumulated transactions received from sensors and actuators grouped in a distinct block by linked nodes organized based on time consensus in a topological order. FlexiChain is the compounded ledger that comprises of multiple independent layer 1 blockchains forming one stronger layer 0 ledger.  One of the independent ledgers in FlexiChain represents the registration and installation process of the nodes. These are virtually mirrored in the FlexiChain by a certain trusted module's public keys and block type. The ledger starts at the same time with the installation step with a predefined number of nodes. 
The ledger is the central record of digital assets through FlexiChain. Existing nodes  are authenticated by new nodes so that the are  authenticated using virtual nodes linked within the network. 

\subsubsection{Trusted Modules} \label{subsec:TPM}

The security hardware is used in this technology for registration. Each module has its predefined keys and one can recognized the other modules. They should be manufactured by a trusted source even though they are just used one time through the whole process until each node can construct new IDs, which eventually will be used as major keys. 

\subsubsection{Block Types} \label{subsec:BlockType}

\begin{figure}[htbp]
	{\includegraphics[width=.80\textwidth,trim={0cm 0cm 0cm 0cm}]{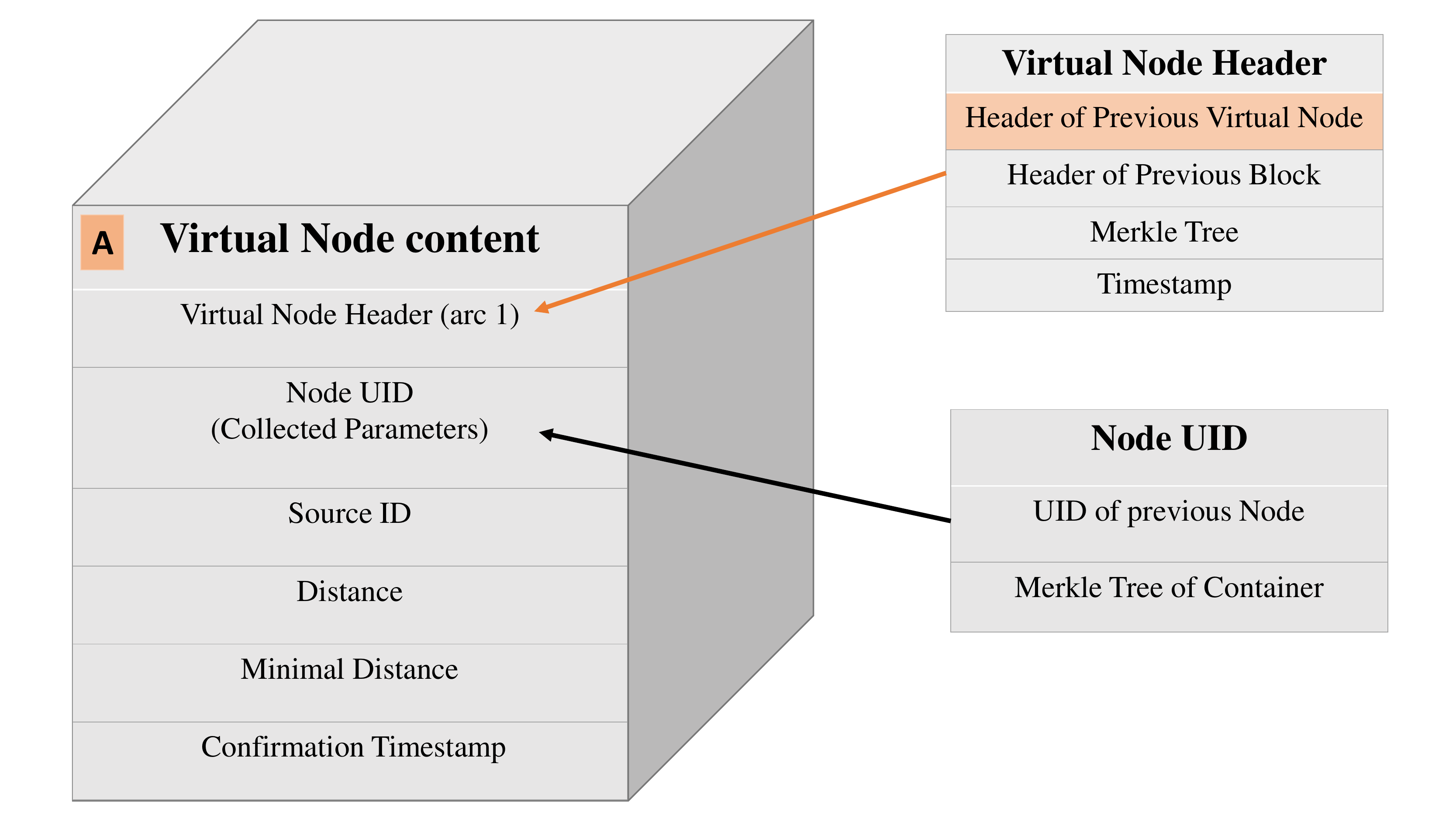}}
	{\includegraphics[width=.80\textwidth,trim={0cm 0cm 0cm 0cm}]{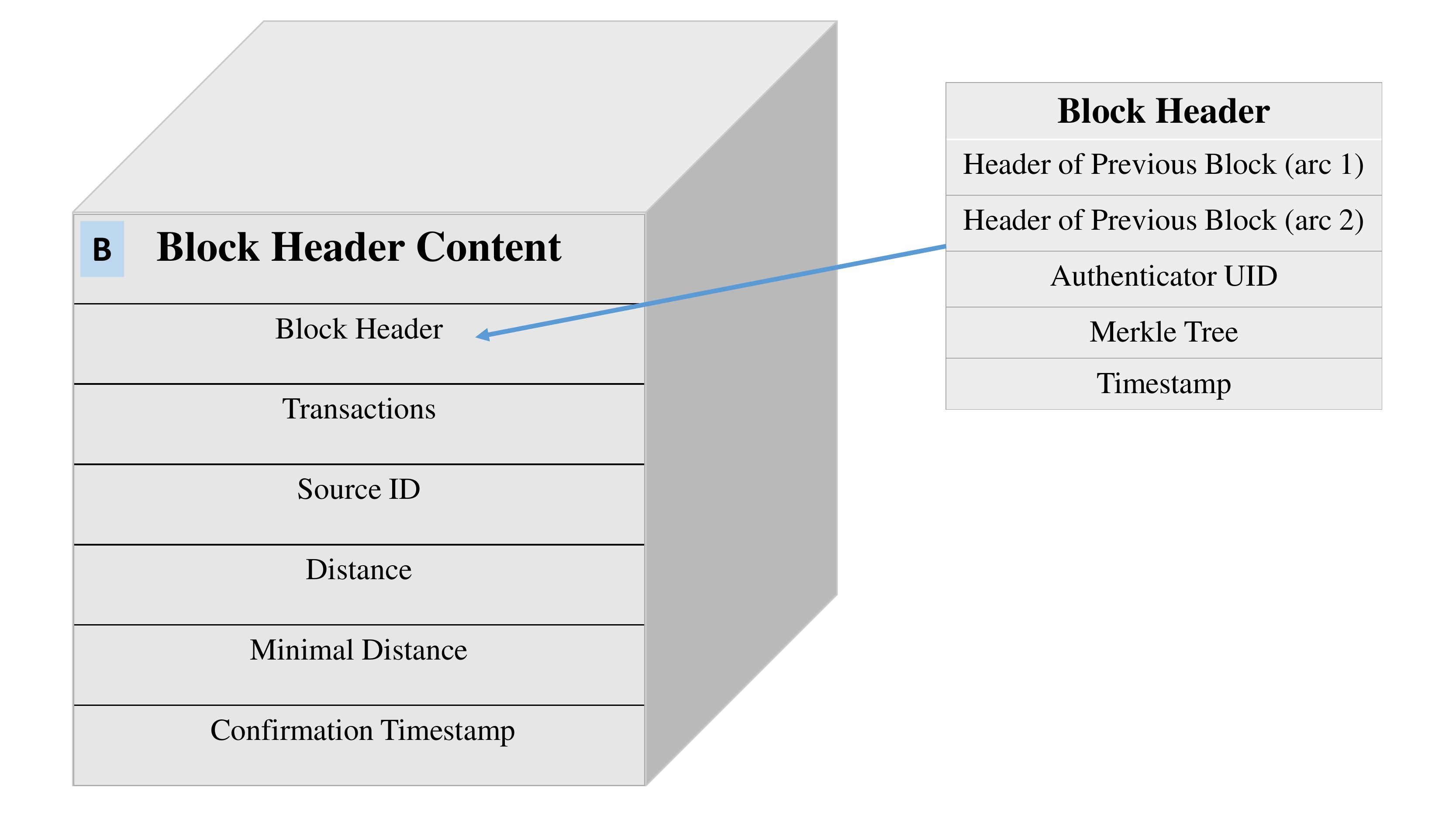}}
	\caption{FlexiChain Block Types}
	\label{FIG:BlockTypes}   
\end{figure}

The types of block used are defined by the digital signatures used for this certain block or by its label. Each distinct type is used for different purposes. Same-type blocks are always linked randomly by one of their arcs. Two types of blocks are used in this implementation: one, which represents the virtual existence of nodes and preserves their UID, and the other type is used for data exchange. A detailed structural representation of these block types in shown in figure \ref{FIG:BlockTypes}.

\subsubsection{NodeChain In FlexiChain}

The proposed method is to provide a decentralized vault that mainly has been proposed to be integrated with FlexiChain. NodeChain is the structure used to create the Unique Identifications (UIDs) based on the node manufacturer specification to form a UID and keep the specification encrypted and stored in the ledger. NodeChain is a series of linked blocks containing UIDs that have been generated using the target hash of extrinsic parameters and rehashed with the previous UID. With each update, and using this method, any changes happening to any node will be perceived by all nodes form the changes in the block header (Virtual Existence) in the network and the Back up Node (BN) can act accordingly. Observing this structure, the similarity of the approach used to monitor the changes in UIDs to the blockchain is clear and has the same objective. NodeChain aims to secure the network by monitoring the changes in block headers using extrinsic parameters that have been extracted from manufacturing features because any changes in the contents, which is the specification, means a change in the block header which gives an indication of the UID which will not match anymore. Also, it reduces the authentication time by decreasing the number of authenticators during the process. The tree of the last UID, once appended to the network, will represent the hash value of all previous UIDs, which means an authentication from the last node is equivalent to all previous nodes. This process is called the Chain of Narration (CN) in which authenticators are linked and listed in each block they authenticate. In bidirectional authentication, the blocks could authenticate the virtual copy of the node and the node will be able to authenticate a virtual node (block) during the operation. In NodeChain the enrollment process is achieved by previously identified hardware such as Trusted Hardware Security that is built based on manufacturer trust.

\subsubsection{NodeChain As An Offline Assisted Distributed Vault}

\begin{figure*}[htpb]
	\centering
	\includegraphics[width=0.90\textwidth,trim={0cm 0cm 0cm 0cm}]{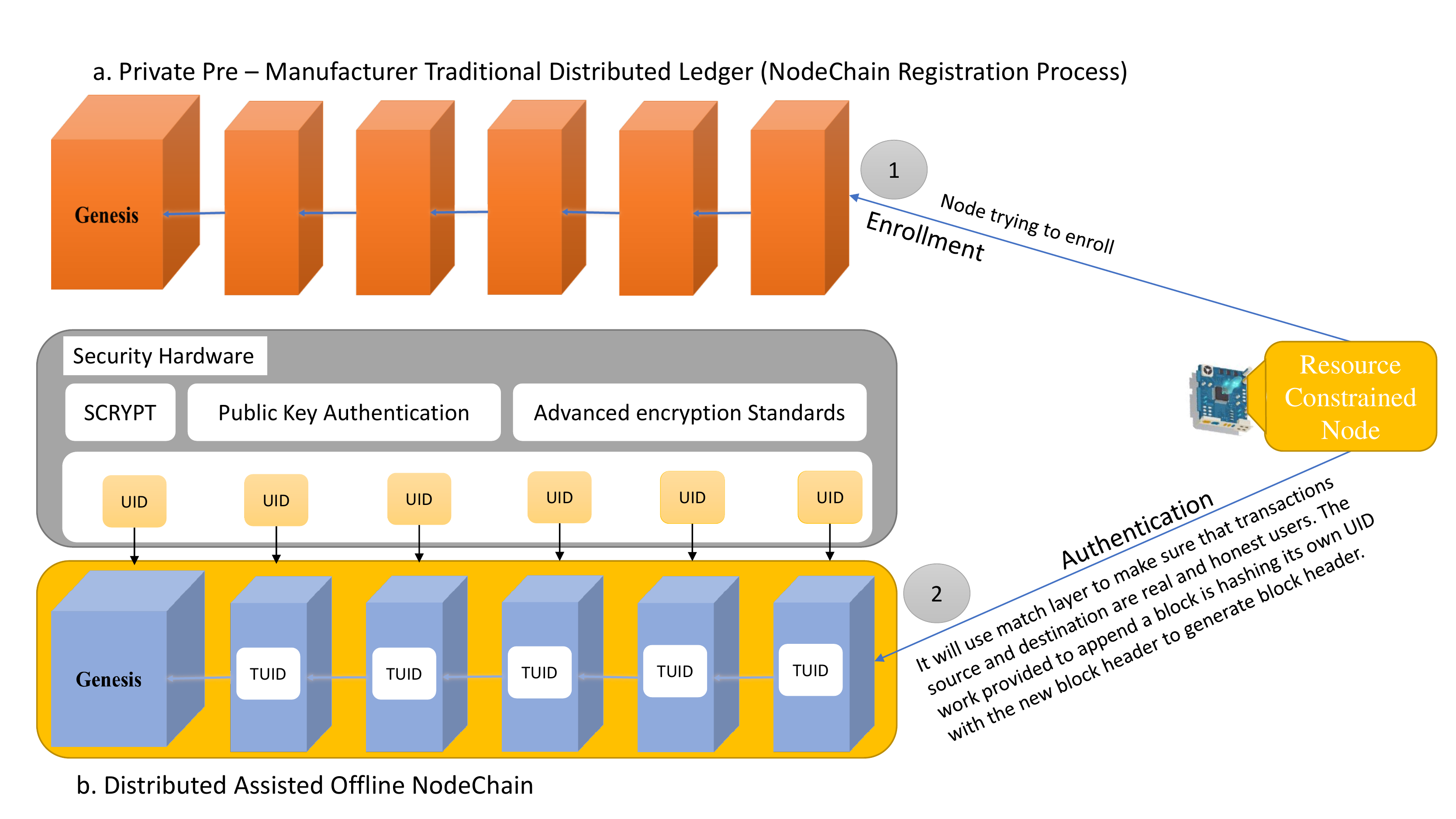}
	\caption{NodeChain As An Offline Assisted Distributed Vault} 
	\label{FIG:AssistedDistributedLedger} 
\end{figure*}

The method is also proposed to be an assisted distributed ledger for other blockchains it order to unify node credentials. If NodeChain is applied to other blockchains, it could reduce latency and could minimize power consumption due to its low computation requirements. In figure \ref{FIG:AssistedDistributedLedger}, a node wants to append a block to the ledger. Each block has multiple transactions and each transaction has a source and destination. In this method, the node first authenticates each transaction using the trusted hardware by comparing TUID with the UID through the match layer between offline ledger and security hardware. The node will create a block and hash its own TUID.  

\subsubsection{Node Types}

There are four types of nodes: 
\begin{enumerate}
	\item The Backup Node (BN) is the cloud of the whole network and the genesis node. The backup node virtual existence in NodeChain is the first block in the ledger. 
	\item Edge Node is a full node that has a Virtual Existence State (VES) and a full ledger. 
	\item Subscriber Node is a node that operate through an edge node. 
	\item Cyber Physical System (CPS) and Internet of Things (IoT) nodes are nodes from which edge nodes collect and exchange their data. An IoT and CPS node could qualify based on the specification acquired since the technology targeting constrained nodes. 
\end{enumerate}
Having a cloud as a node and a set of edge nodes which share the same capabilities is very helpful to avoid Single Point Failure (SPF) Attack. The backup node is the network server from which all nodes will start their Peer to Peer (P2P) communication. Other than keeping the ledger state synced, the backup node will operate when there are no transaction to maintain the ledger state. Back up node and edge nodes all have the same authorities over the network. Other nodes might be qualified for being a full node based on the specification of each node.

\subsubsection{Authentication} \label{subsubsec:Authentication}

\begin{figure*}[htbp]
	\centering
	\subfigure[Node Authentication Requirements]
	{\includegraphics[width=0.75\textwidth,trim={0cm 0cm 0cm 0cm}]{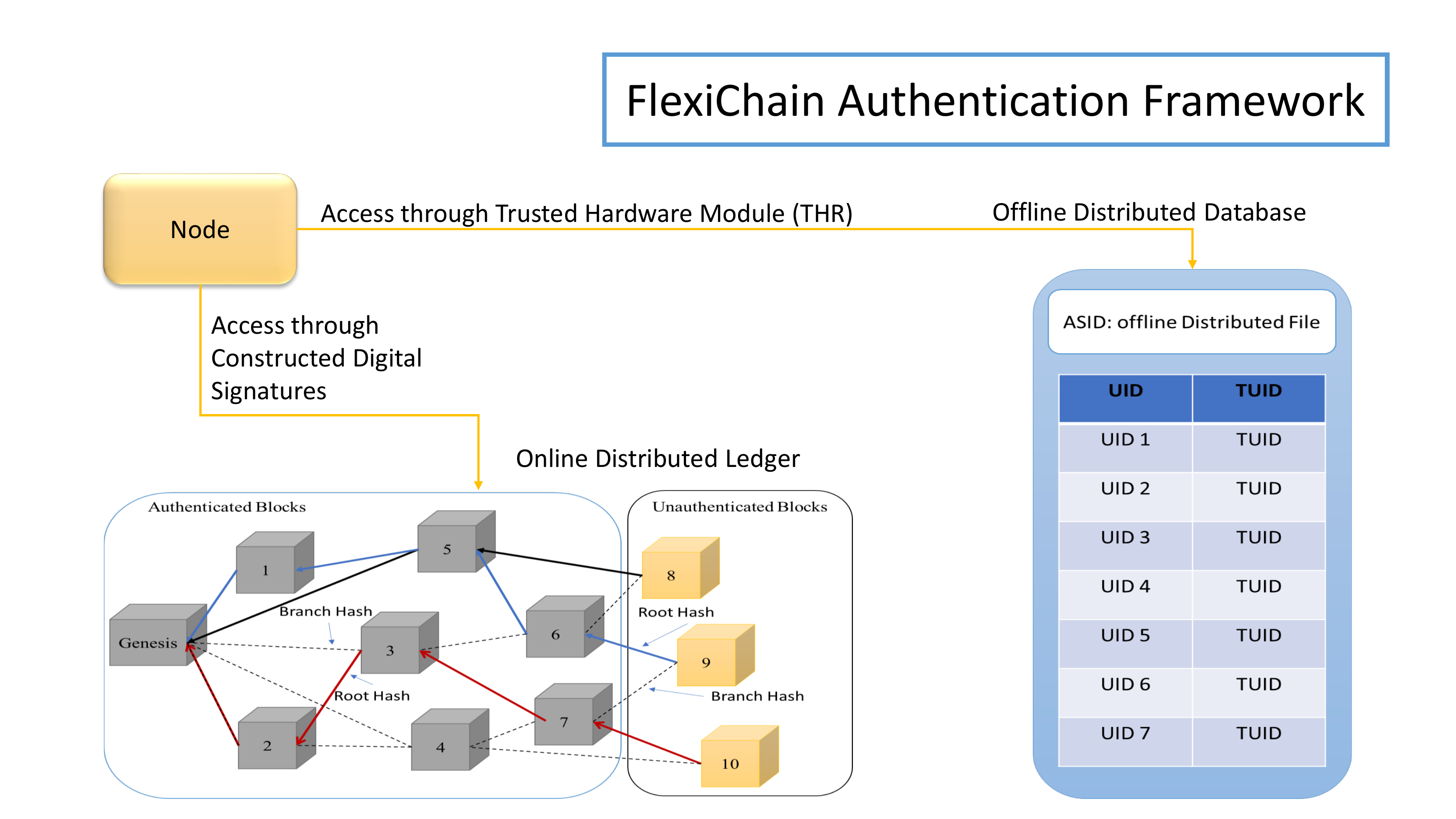}}
	\subfigure[Node Authentication Process]
	{\includegraphics[width=0.80\textwidth,trim={0cm 0cm 0cm 0cm}]{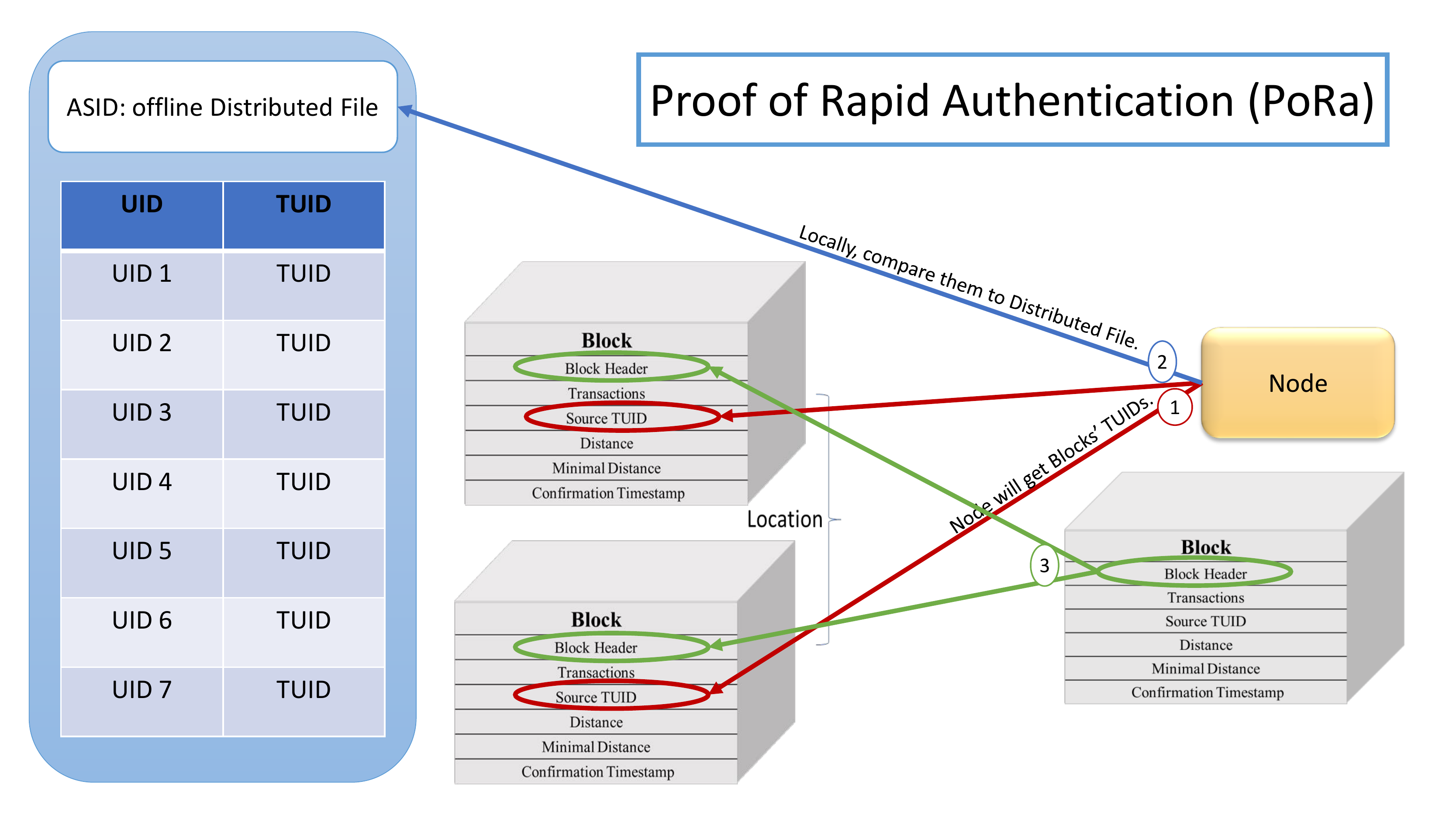}}
	\subfigure[Node Block Appending]
	{\includegraphics[width=0.70\textwidth,trim={0cm 0cm 0cm 0cm}]{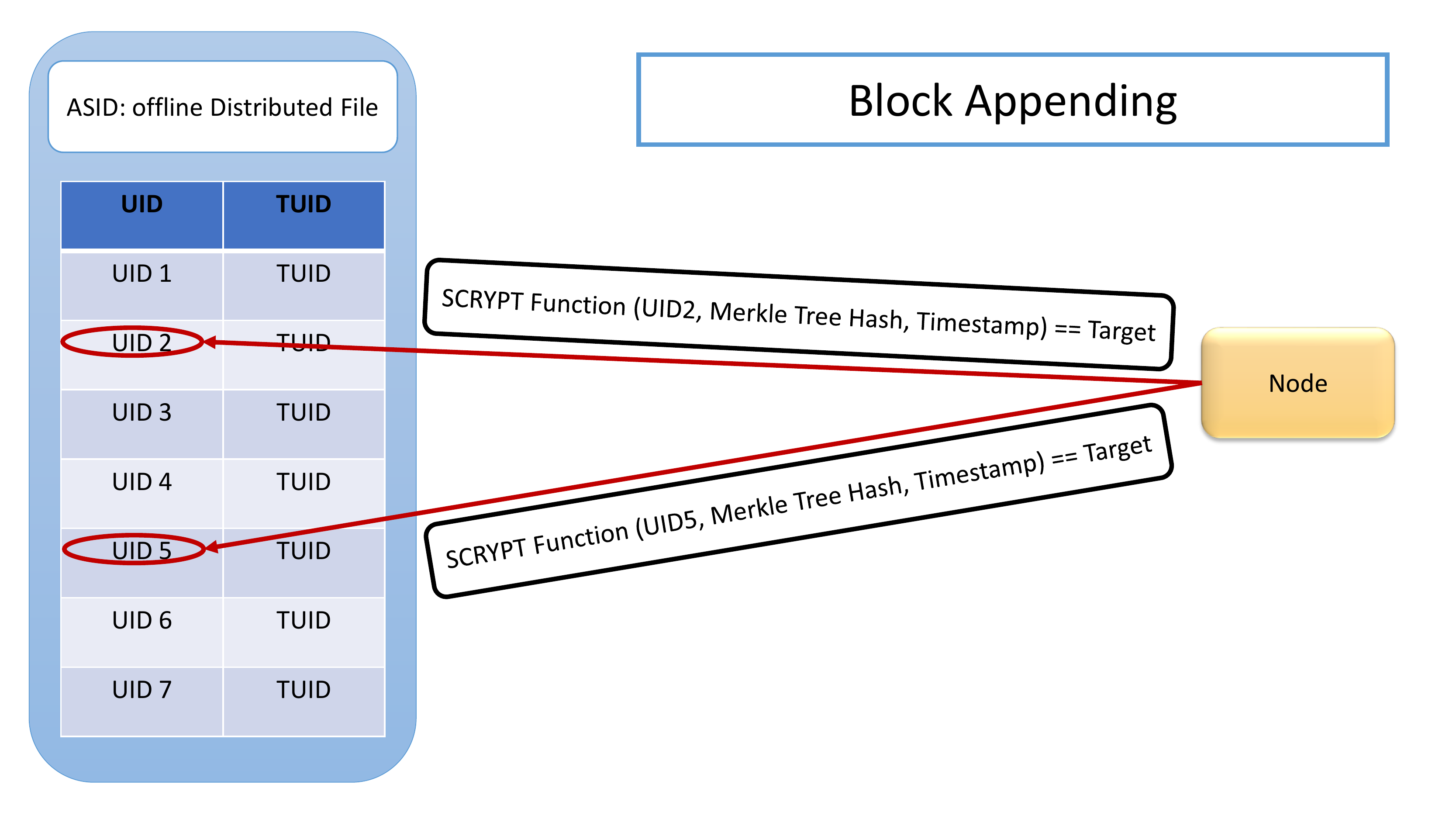}}
	\caption{Proof of Rapid Authentication Framework}
	\label{FIG:ProofofRapid_Authentication_Framework}   
\end{figure*}

Referring to figure \ref{FIG:ProofofRapid_Authentication_Framework}, node C of the network sends transactions of type B $\hookrightarrow BTrxs$. Similarly, they will be collected based on sender, type, time consensus and Type of block $BTrxs (n) \hookrightarrow Bb $. A new header is formed by the authenticator UID, and two other hashes. Once a block receives confirmations, the authenticators' UID $AUIDS$ will be listed until $AUIDS == All UIDs$. 

\subsection{Proposed Method}
\label{subsec:Algorithm Discussion}

\subsubsection{Registration}
\label{subsubsec:Registrationchap5}

In this section, the process of making nodes recognize each other in the network through NodeChain will be explained. This is the enrollment process which each node should perform once when they want to join the network. This process is assumed to be private prior to the use in public FlexiChain or any other DLT.

In this process, hardware security keys can recognize each other through their key signatures that are predefined to them. Nodes recognize each other through the construct signatures. The recognition between hardware security is a manufacturer trust. Nodes pre-recognition takes place through an enrollment private process. 

\subsubsection{Distributed Accessible Unique Identification NodeChain}

\begin{figure*}[htbp]
	\centering
	\subfigure[Node Initial Request]
	{\includegraphics[width=0.70\textwidth,trim={0cm 0cm 0cm 0cm}]{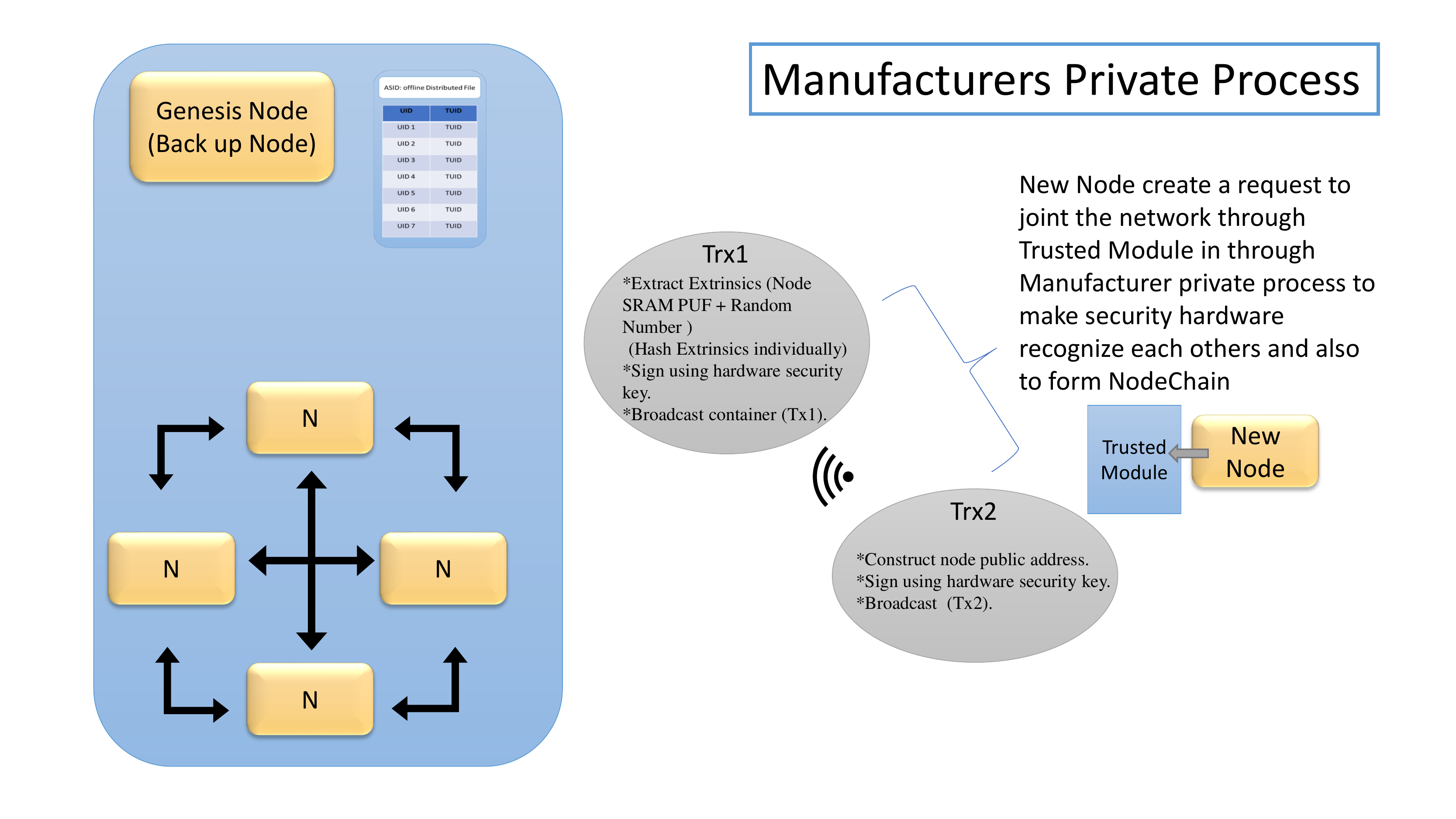}}
	\subfigure[Network Operation and Response]
	{\includegraphics[width=0.80\textwidth,trim={0cm 0cm 0cm 0cm}]{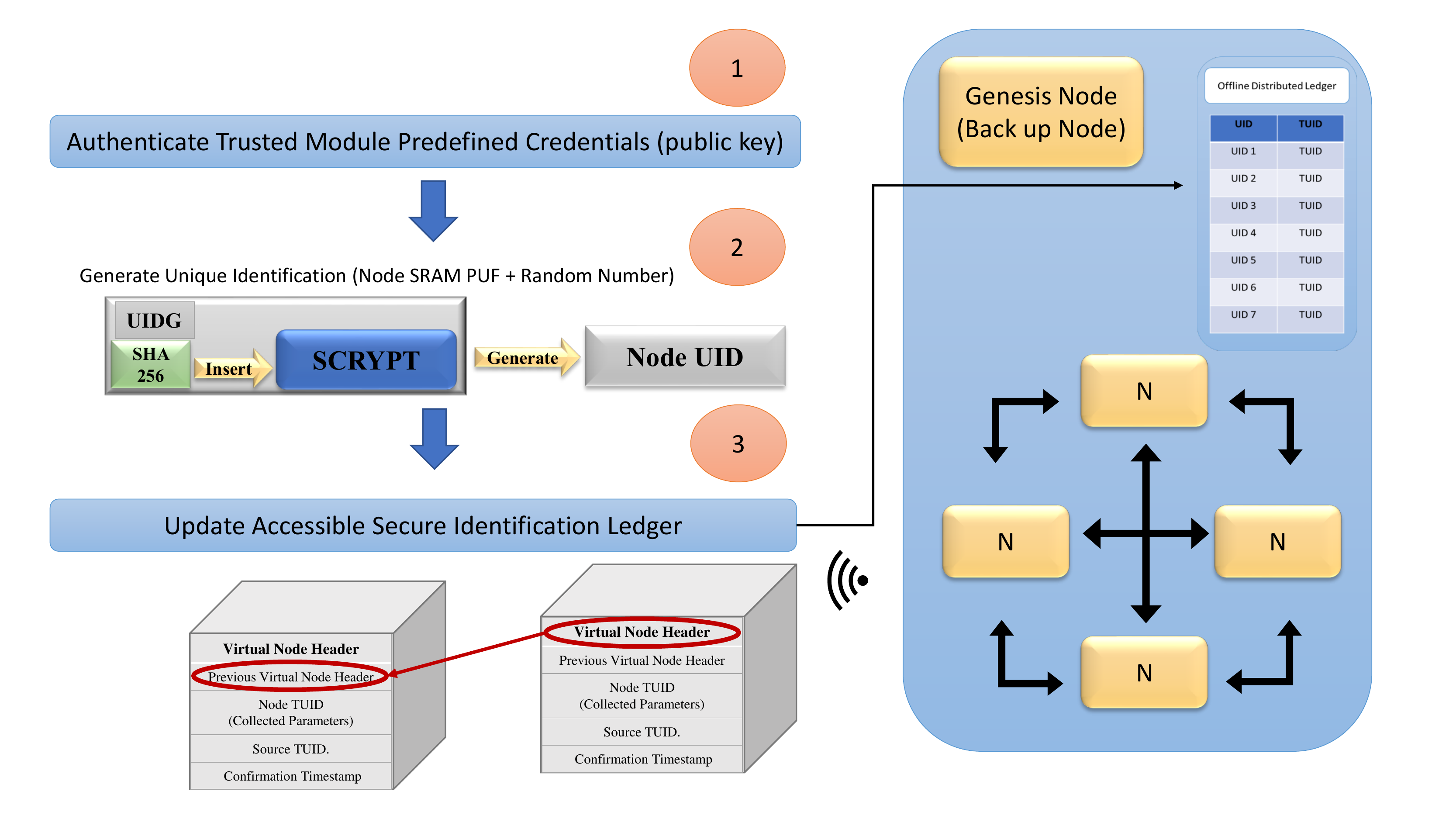}}
	\subfigure[Node Part of The Network]
	{\includegraphics[width=0.70\textwidth,trim={0cm 0cm 0cm 0cm}]{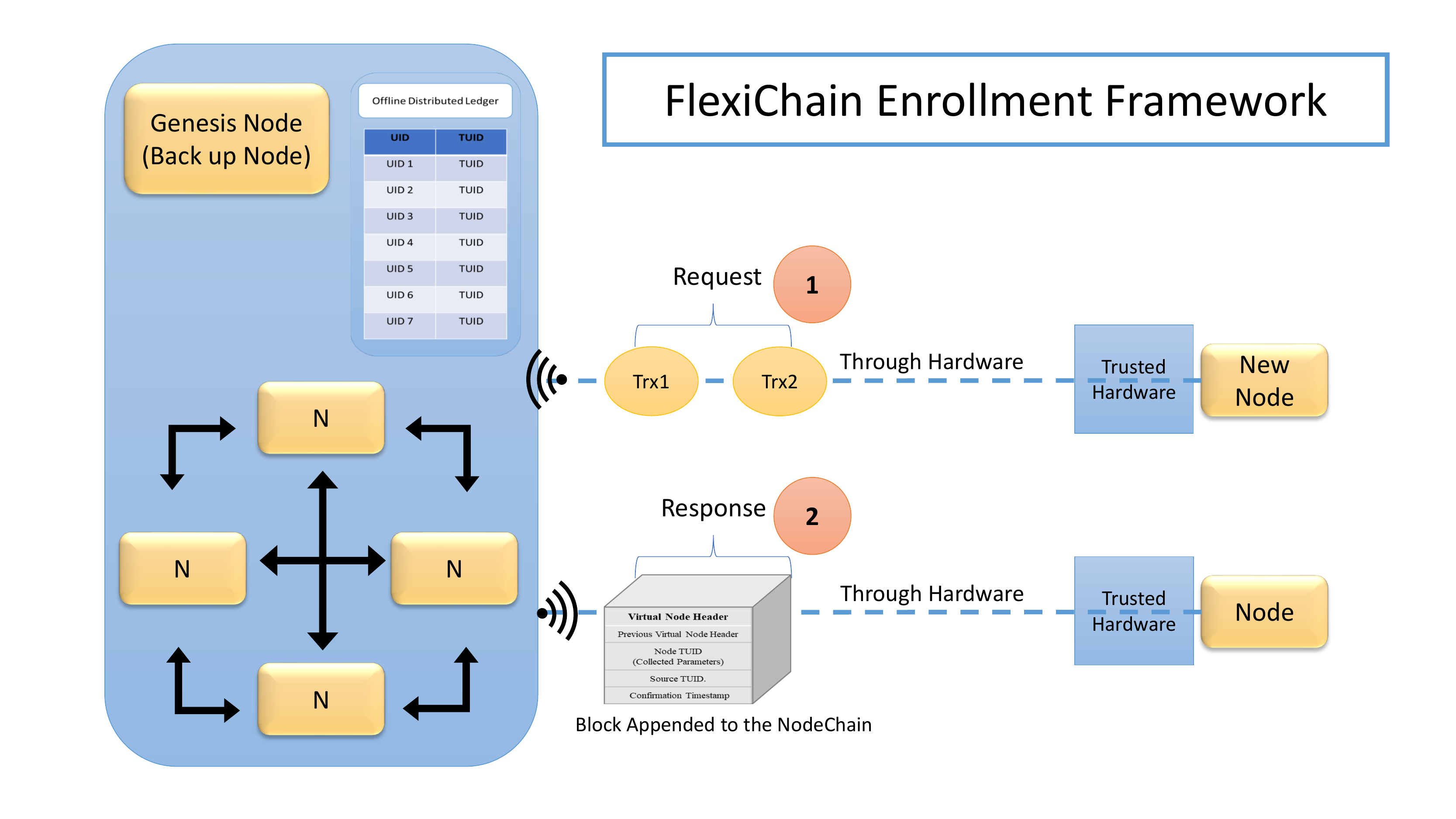}}
	\caption{Proof of Rapid Authentication Framework}
	\label{FIG:ProofofRapid_Registration_Framework}   
\end{figure*}

The initial stage of the network starts with only one edge node which is the back up node. The virtual copy of this node is the genesis block of the layer 0 ledger. Prior to initiating the network, a certain number of Trusted Hardware Modules (TPM) are predefined to the backup node by their public key in the distributed vault. This identification will create a secure channel for nodes to join the network and construct their own signatures using TPM. This process involves multiple steps and components. In figure \ref{FIG:ProofofRapid_Registration_Framework}(a), the first step is that the joining node will communicate with the first and genesis node in the network through TPM to grant a virtual copy in the ledger. The node will broadcast two transactions called a ``Request'', each of which reveal certain parameters called extrinsic parameters. The back up node receives transactions and generate a unique UID using Unique Identification Generator (UIDG) and updates the distributed file. The back node generates a tokenized version of the UID and creates a virtual copy (block) add block to the ledger and updates the ledger. This step is called ``Response''. Now, the joining node is part of the network and could get a copy of ledger and distributed file. 

\subsubsection{Extrinsic Parameters}

Parameters are used in this method to create a unique identity derived from two sources: the real manufacturing identity and a constructed identity \cite{Safeguard}. Manufacturing Identity parameters are MAC address, firmware parameters, SRAM-PUF signature, process power parameters and constructed identity such as location parameters. IP address, and constructed public ID, are all examples of extrinsic parameters. The type and the number of extrinsic parameter choice will be based on the application. The extrinsic parameters are all integrated in one UID (signature). This signature is the result feeding this sequence of values to a SCRYPT function that would generate 126 key length that represent the actual UID. The parameters extracted will be the first indicator for accepting a node to be part of the network. Extrinsic parameter are transacted as two containers: first, is the container that has the parameter fit the application targeted. Second, is the Constructed Identity (CI) which is the second digital signature used in this technology. The predefined is the one used for layer zero, and the constructed one used for layer one.

\subsubsection{Request and Response}

\begin{figure*}[htbp]
	\centering
	\includegraphics[width=0.90\textwidth,trim={0cm 0cm 0cm 0cm}]{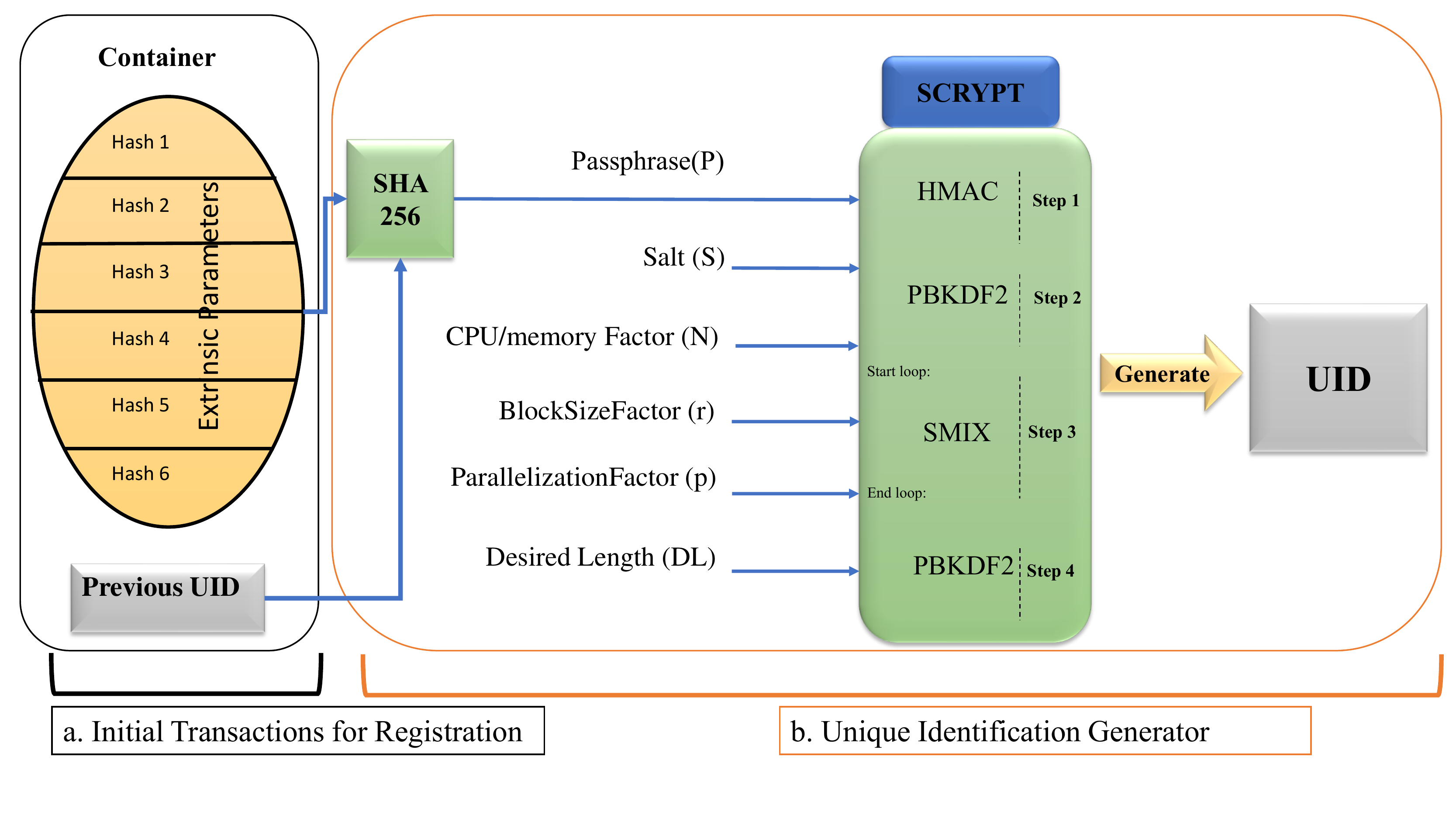}
	\caption{The Registration Process in FlexiChain} 
	\label{FIG:Images_UIDG_PoRa} 
\end{figure*}

In any network, to establish a peer to peer communication is an essential part of a Distributed Ledger Technology (DLT). FlexiChain relies on this type of communication. The idea behind this operation is to make two nodes communicate directly after they established a secure channel among them. In addition to the communication channel, in FlexiChain, there is a security channel that guarantees the integrity of the nodes by sharing nodes' specification and keep it part of the ledger. This process is accomplished through requests and responses. Prior to enrollment, the process uses Trusted Hardware Security to initialize the first communication with previous node to fulfill enrollment requirements. A trusted hardware security has its own keys and a prior definition to backup node which gives the new node the opportunity to communicate using the Trusted Hardware Security credentials and to start constructing its own during the process. First step in this operation is starting to extract Extrinsic Parameters. The joining node will start to create it own containers of data and share it with a previous node that is eligible to create a Unique Identification (UID) through a request which is: transactions broadcast by devices to the network and received by the Back up Node (BN) or any other qualified Edge Node (EN) to append the node to the NodeChain. Two transactions (containers) are generated. First, include the hash of extrinsic parameters. Second, the constructed public ID. The response is the block broadcast to the whole network as the virtual existence of the same node that includes the UID and the constructed public key. The request and response operation is depicted in figure \ref{FIG:ProofofRapid_Registration_Framework}. The response is only generated by BN or EN. Figure \ref{FIG:Images_UIDG_PoRa} presents the content of the requests.  

\subsubsection{Unique Identification Generator (UIDG)}

The UIDG function initially exists only in the backup node. Once an edge node joins and runs the virtual node, the edge node will be able to execute UIDG for other nodes. This function is used to generate the UID for each node. The UIDG includes two modules: the first one is the SCRYPT function \cite{SCRYPT}, and the second is the SHA256 function. The extrinsic parameters and previous UID will be fed to the SHA256 module and the result will be fed to the SCRYPT function. Figure \ref{FIG:Images_UIDG_PoRa} presents modules associated with the UIDG. The scrypt function parameters are used but are not supposed to be shared among users.

\subsubsection{SCRYPT Function} 

This hash function uses multiple factors in addition to the value needed to be hashed to generate a hash value, such as salt, CPU or memory, mixing loops and parallelization to increase the difficulty in the event of an attack. Parallelization is a way to double the processing power needed just by changing the factor from 1 to 2. The factors of SCRYPT is not to be shared among users and should be secret to the network \cite{SCRYPT}.

\subsubsection{Virtual Existence State (VES)}

\begin{figure*}[htbp]
	\centering
	\includegraphics[width=0.90\textwidth,trim={0cm 0cm 0cm 0cm}]{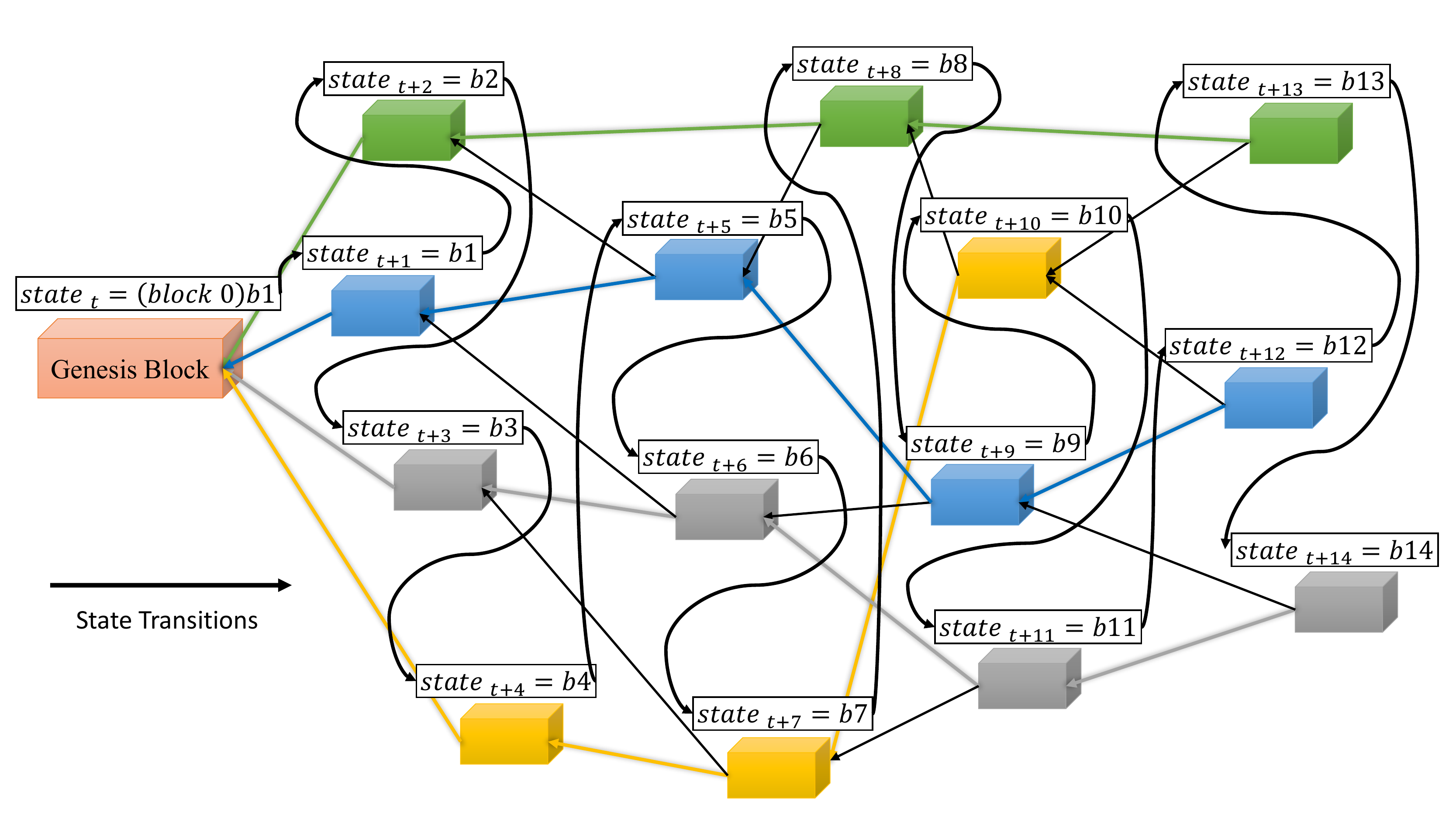}
	\caption{State Transition for VES} 
	\label{FIG:StateTransition} 
\end{figure*}

In Ethereum Virtual Machine (EVM), the machine keeps its state synced among all participants. The state machine transition is based on the changes of the ledger. Similarly, the Virtual Existence State, is a changeable state that represents a state at a certain time. This approach is essential to keep the same state among participants and to keep their version of the ledger updated. Also, when a new node is joining the network, the node will be able to download and track the ledger version from the state value. Figure \ref{FIG:StateTransition} presents the transition state for all nodes including backup node. The transition state changes with every new block joining the ledger. This ledger should be maintained with all participant. 

\section{Experimental Results}
\label{sec:Implementation}

\subsection{Security And Privacy Analysis}

FlexiChain takes into account both software and hardware and software security. Various security vulnerabilities can be used to test the technology \cite{Blockchain_Based_DeepLearning_CPS} such as phishing attack, Sybil attack, brute force attack, DDOS attack, and identity attack.

\subsection{Category Attack-1 (Sybil)}

Sybil attacks are set up by giving the same node many credentials. Each request in a blockchain network is broadcast to many nodes, and there is no central authority to verify their legitimacy. A hacker launches a Sybil attack when he or she gains control of many network nodes at once. Once the victim is encircled by malicious nodes, all of their transactions are frozen. At this point, the victim is completely vulnerable to double-spending schemes. Despite the difficulty in detecting and preventing a Sybil attack, several solutions have proven effective. These include increasing the cost of generating a new identity, demanding trust before joining the network, and assigning user authority based on reputation. This attack category is described as a node launching an attack to control other nodes. To control a node, credentials should be known.  

\subsection{Category Attack-2 (Phishing)}

In 2018, an online seed generator by the name of iotaseed.io (which is no longer accessible) was used to launch an attack on IOTA wallets. The service was used in a phishing campaign, and logs containing secret seeds were obtained by the hackers. Consequently, in January of 2018, hackers successfully stole over 4 million dollars worth of IOTA from victims' wallets. The first step in a phishing attempt is typically the delivery of a bulk email or message to anyone who might fall for the scam. Sometimes it can even appear to be from a reliable source. This attack category is described as a node identity has been stolen and been used to compromise the data or even to launch a Distributed Denial Of Service attack (DDOS).

\subsection{Category Attack-3 (51\%)}

Gaining control of 51\% of the network's hash rate allows an attacker to launch a majority attack and build a new fork that eventually supersedes all previous forks. This attack was the first and only one that had been identified as a weakness in the blockchain and had seemed highly improbable. However, 51\% attacks have been successful against at least five cryptocurrencies so far: Verge, ZenCash, Monacoin, Bitcoin Gold, and Litecoin Cash. Each time, the hackers amassed enough hashing power to break into the system and steal millions. To put it simply, the August 2020 51\% attack on Ethereum Classic (ETC) caused about 5.6 million dollars worth of ETC to be double-spent. Evidently well-versed in the ETC protocol, the attacker mined 4,280 blocks over the course of four days before the platform detected an attack. The next 51\% attack on ETC occurred only five days later, when a miner reorganized the network by removing 4,000 blocks. Observing the above definition of this attack, this attack can take place in networks that use Proof of Stake as well. Assuming that only 5 staking pools have the majority of a certain currency, they have applied the attack and controlled the network without the need of performing an actual technological attack and that due to the authority a node gains while having a high stake. This attack category is described as a group of pools intend to control 51\% of network resources. 

\subsection{Category Attack-4 (Brute force)}

Cryptocurrencies' private keys are extremely difficult to brute-force with modern technologies. The advent of quantum computers, which can process data at an exponentially faster rate than traditional computers, may, however, make such attacks feasible. While quantum computers are still a few years from widespread use, they nevertheless present a potential threat to digital currencies in the near future. This type of attack has been considered as a threat and described as a successful attack has been performed and has gotten the private key of a certain node to send a compromised data or act maliciously inside the network.

Simulation has been performed on values of $\alpha\kappa$, $\beta\kappa$, $\overline{\alpha\kappa}$, $\overline{\beta\kappa}$, $\phi\kappa$ and $\rho\kappa$ are assigned values between (0.9-1) based on the possibility and difficulty of a particular attack. The number of nodes are chosen to be (4 - 64) which represents $n$ in the equation. For an attack that has been experienced similar to examples given attack categories to the best of our knowledge will get lower difficulty and for those without an experienced attack will get higher difficulty \cite{Blockchain_Based_DeepLearning_CPS}.

\subsubsection{Blockchain}

The total probability of the aforementioned attacks to take place in a public ledger (Blockchain technology) either using Proof of Work or Proof of Stake is represented as ($P(BC)$) and has 4 parts each of which represents an attack category: 
\begin{equation} 
	P(BC) = P(BC-1) + P(BC-2) + P(BC-3) + P(BC-4)
	\label{eq:BC_eq1} 
\end{equation}
$P(BC-1)$ represents category attack 1 and can be done by first a successful attack over node, and ability to get the nodes' credentials. The probability can be calculated by 
\begin{equation} 
	P(BC-1)= \frac{1}{4}\prod_{a}^{n}\alpha\kappa\times\frac{1}{4}\prod_{a}^{n}\beta\kappa  
	\label{eq:BC_eq2} 
\end{equation}
where $\alpha\kappa$ is the probability of performing a successful attack over a node, and $\beta\kappa$ the probability of stealing credentials. The probability is calculated for each node in the case of Bitcoin and Ethereum by multiplying all nodes' probability of full control ($\alpha\kappa a$-$\alpha\kappa n$) and credential stealing ($\beta\kappa a$ to $\beta\kappa n$). 

$P(BC-2)$ represents category attack 2 and can be done by first a successful attack over node, and ability to get the nodes' credentials. The probability can be calculated by
\begin{equation} 
	P(BC-2)= \frac{1}{4}\prod_{a}^{n}\alpha\kappa\times\frac{1}{4}\prod_{a}^{n}\beta\kappa  
	\label{eq:BC_eq3} 
\end{equation}
$P(BC-3)$ represents category attack 3 and can be done by first a successful attack over node, and ability to get the nodes' credentials. The probability can be calculated by
\begin{equation} 
	P(BC-3)= \frac{1}{4}\prod_{a}^{n}\alpha\kappa\times\frac{1}{4}\prod_{a}^{n}\beta\kappa  
	\label{eq:BC_eq4} 
\end{equation}
$P(BC-4)$ represents category attack 4 and can be done by first a successful attack over node, and ability to get the nodes' credentials. The probability can be calculated by
\begin{equation} 
	P(BC-4)= \frac{1}{4}\prod_{a}^{n}\alpha\kappa\times\frac{1}{4}\prod_{a}^{n}\beta\kappa  
	\label{eq:BC_eq5} 
\end{equation}
Using these equations, the total probability is obtained:
\begin{equation} 
	P(BC)= \frac{1}{4}\prod_{a}^{n}\alpha\kappa\times\frac{1}{4}\prod_{a}^{n}\beta\kappa+\frac{1}{4}\prod_{a}^{n}\alpha\kappa\times\frac{1}{4}\prod_{a}^{n}\beta\kappa+\frac{1}{4}\prod_{a}^{n}\alpha\kappa\times\frac{1}{4}\prod_{a}^{n}\beta\kappa+\frac{1}{4}\prod_{a}^{n}\alpha\kappa\times\frac{1}{4}\prod_{a}^{n}\beta\kappa  
	\label{eq:BC_eq6} 
\end{equation}

\begin{figure}[htpb]
\centering
\includegraphics[width=0.90\textwidth]{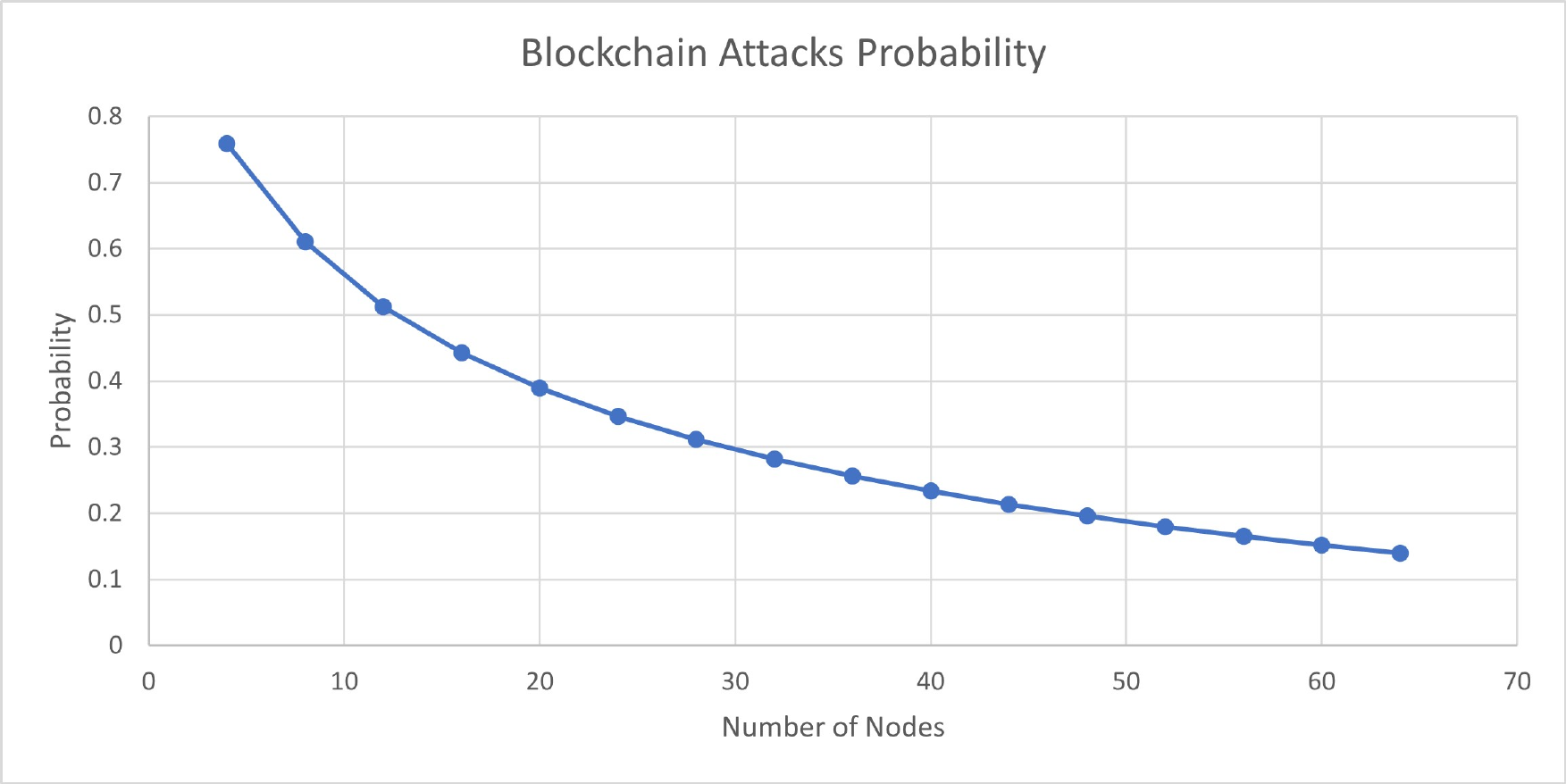}\hspace{1mm}%
\caption{Results for 4 Categories of Attack Scenarios in the Blockchain.}
\label{fig:ResultsBlockchainSecurity}
\end{figure}

The probabilities obtained from our security analysis as a function of the number of authenticated nodes are shown in figure \ref{fig:ResultsBlockchainSecurity} and summarized in Table \ref{tab:BlockSecTable}.

\begin{table}[htpb]
	\centering
	\caption{Probabilities for four categories of attack in the Blockchain}
	\label{tab:BlockSecTable}
\footnotesize
	\begin{tabular}{p{1cm} p{1.65cm} p{1.65cm} p{1.65cm} p{1.65cm} l}
		\toprule
		Number of Nodes & Attack-1 Probabilities  & Attack-2 Probabilities & Attack-3 Probabilities  & Attack-4 Probabilities & Summation\\
		\midrule
		4     & 0.180269132 & 0.230686174 & 0.230686174 & 0.117563132 & 0.759204611 \\
		24    & 0.035142136 & 0.154322535 & 0.154322535 & 0.002703503 & 0.34649071 \\
		44    & 0.0068507 & 0.103237418 & 0.103237418 & 6.21702E-05 & 0.213387706 \\
		64    & 0.001335493 & 0.069062917 & 0.069062917 & 1.42968E-06 & 0.139462757 \\
		\bottomrule
	\end{tabular}
\end{table}

\subsubsection{FlexiChain With NodeChain}

On the other hand, FlexiChain technology supported by NodeChain has its own extra security layers. The total probability of the aforementioned attacks to take place in a public ledger (FlexiChain technology) using NodeChain following Upgrade Proof of Rapid Authentication as a consensus is represented as ($P(FC)$) and it has 4 parts, each of which represents an attack category:
\begin{equation} 
	P(FC) = P(FC-1) + P(FC-2) + P(FC-3) + P(FC-4)
	\label{eq:FC_eq7} 
\end{equation}

$P(FC-1)$ represents category attack 1 and can be done by first a successful attack over a node, and ability to get the nodes' credentials, trusted attached hardware, and Unique Identification (UID). The probability can be calculated by 
\begin{equation} 
	P(FC-1)= \frac{1}{4}\prod_{a}^{n}\overline{\alpha\kappa}\times\frac{1}{4}\prod_{a}^{n}\overline{\beta\kappa}
	\times\frac{1}{4}\prod_{a}^{n}\phi\kappa\times\frac{1}{4}\prod_{a}^{n}\rho\kappa
	\label{eq:FC_eq8} 
\end{equation}
where $\overline{\alpha\kappa}$ is the probability of performing a successful attack over a node, and $\overline{\beta\kappa}$ the probability of stealing credentials. $\phi\kappa$ is the probability to acquire a trusted module private key. $\rho\kappa$ is the probability to gain access to NodeChain and obtain Tokenized UID and access the offline distributed vault to match with a real UID. The probability is calculated for each node $a$ in the case of FlexiChain by multiplying all nodes' probability of full control ($\alpha\kappa a$-$\alpha\kappa n$) and credentials stealing ($\beta\kappa a$ to $\beta\kappa n$). 

$P(FC-2)$ represents category attack 2 and can be done by first a successful attack over node, and ability to get the nodes' credentials. The probability can be calculated by
\begin{equation} 
	P(FC-2)= \frac{1}{4}\prod_{a}^{n}\overline{\alpha\kappa}\times\frac{1}{4}\prod_{a}^{n}\overline{\beta\kappa}
	\times\frac{1}{4}\prod_{a}^{n}\phi\kappa\times\frac{1}{4}\prod_{a}^{n}\rho\kappa 
	\label{eq:FC_eq9} 
\end{equation}

$P(FC-3)$ represents category attack 3 and can be done by first a successful attack over node, and ability to get the nodes' credentials. The probability can be calculated by
\begin{equation} 
	P(FC-3)= \frac{1}{4}\prod_{a}^{n}\overline{\alpha\kappa}\times\frac{1}{4}\prod_{a}^{n}\overline{\beta\kappa}
	\times\frac{1}{4}\prod_{a}^{n}\phi\kappa\times\frac{1}{4}\prod_{a}^{n}\rho\kappa 
	\label{eq:FC_eq10} 
\end{equation}

$P(FC-4)$ represents category attack 4 and can be done by first a successful attack over node, and ability to get the nodes' credentials. The probability can be calculated by
\begin{equation} 
	P(FC-4)= \frac{1}{4}\prod_{a}^{n}\overline{\alpha\kappa}\times\frac{1}{4}\prod_{a}^{n}\overline{\beta\kappa}
	\times\frac{1}{4}\prod_{a}^{n}\phi\kappa\times\frac{1}{4}\prod_{a}^{n}\rho\kappa 
	\label{eq:FC_eq11} 
\end{equation}

Using these equations, the total probability is obtained:

\begin{equation}
\begin{aligned}
	P(FC)= \frac{1}{4}\prod_{a}^{n}\overline{\alpha\kappa}\times\frac{1}{4}\prod_{a}^{n}\overline{\beta\kappa}
	\times\frac{1}{4}\prod_{a}^{n}\phi\kappa\times\frac{1}{4}\prod_{a}^{n}\rho\kappa\\+ \frac{1}{4}\prod_{a}^{n}\overline{\alpha\kappa}\times\frac{1}{4}\prod_{a}^{n}\overline{\beta\kappa}
	\times\frac{1}{4}\prod_{a}^{n}\phi\kappa\times\frac{1}{4}\prod_{a}^{n}\rho\kappa\\+ \frac{1}{4}\prod_{a}^{n}\overline{\alpha\kappa}\times\frac{1}{4}\prod_{a}^{n}\overline{\beta\kappa}
	\times\frac{1}{4}\prod_{a}^{n}\phi\kappa\times\frac{1}{4}\prod_{a}^{n}\rho\kappa\\+ \frac{1}{4}\prod_{a}^{n}\overline{\alpha\kappa}\times\frac{1}{4}\prod_{a}^{n}\overline{\beta\kappa}
	\times\frac{1}{4}\prod_{a}^{n}\phi\kappa\times\frac{1}{4}\prod_{a}^{n}\rho\kappa\\+ \frac{1}{4}\prod_{a}^{n}\overline{\alpha\kappa}\times\frac{1}{4}\prod_{a}^{n}\overline{\beta\kappa}
	\times\frac{1}{4}\prod_{a}^{n}\phi\kappa\times\frac{1}{4}\prod_{a}^{n}\rho\kappa  
	\label{eq:FC_eq12} 
\end{aligned}
\end{equation} 

\begin{figure}[htpb]
	\centering
	\includegraphics[width=0.90\textwidth]{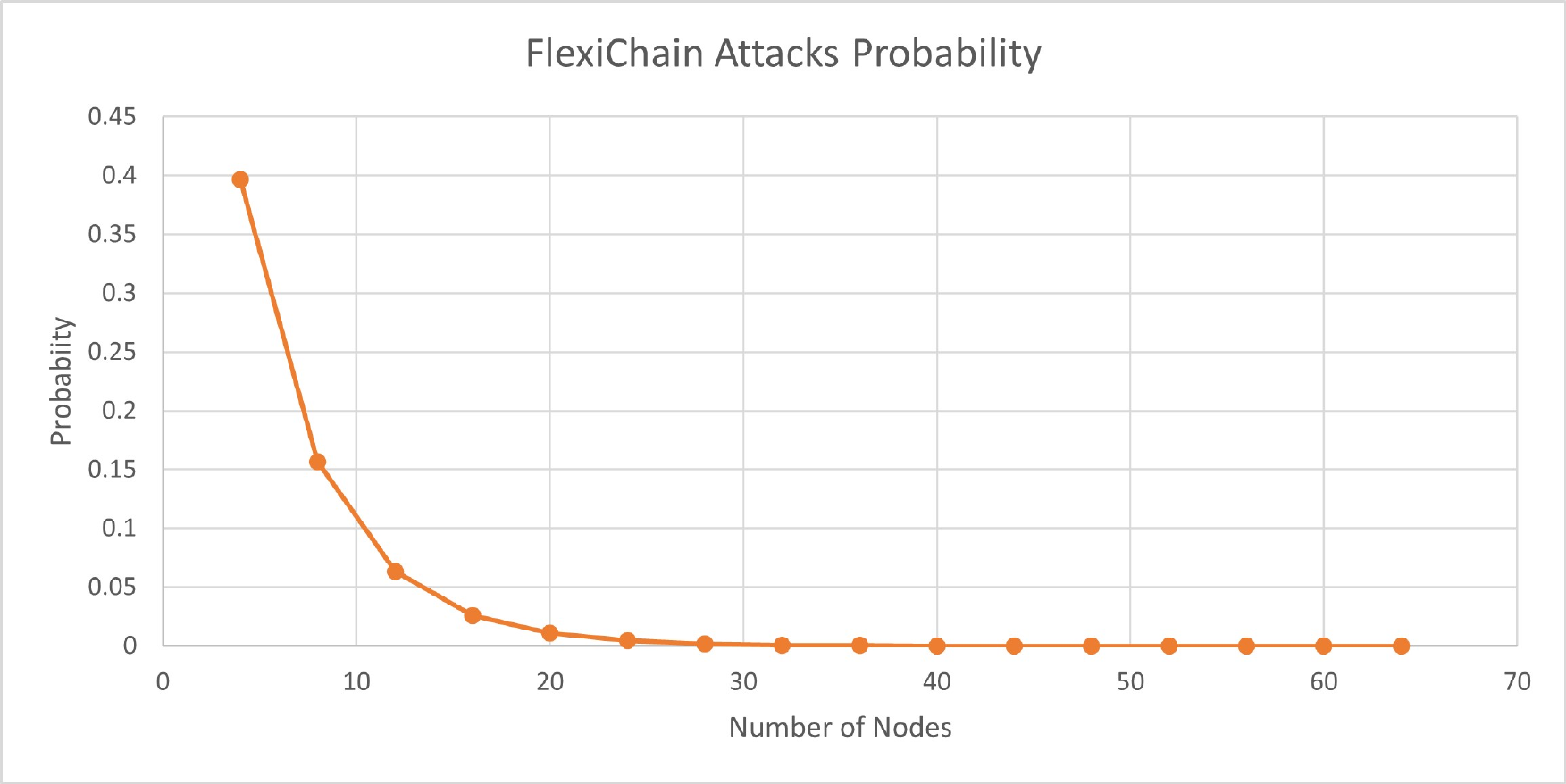}\hspace{1mm}%
	\caption{Probabilities for four categories of attack in FlexiChain}
	\label{fig:ResultsFlexiChainSecurity}
\end{figure}

The probabilities obtained from our security analysis as a function of the number of authenticated nodes are shown in figure \ref{fig:ResultsFlexiChainSecurity} and summarized in Table \ref{tab:FlexiSecTable}.

\begin{table}[htpb]
	\centering
	\caption{Probabilities for four categories of attack in FlexiChain}
	\label{tab:FlexiSecTable}
\footnotesize
	\begin{tabular}{p{1cm} p{1.65cm} p{1.65cm} p{1.65cm} p{1.65cm} l}
		\toprule
		Number of Nodes & Attack-1 Probabilities  & Attack-2 Probabilities & Attack-3 Probabilities  & Attack-4 Probabilities & Summation\\
		\midrule
		4     & 0.104995786 & 0.111672675 & 0.111672675 & 0.068473361 & 0.396814497\\
		24    & 0.001371928 & 0.001459171 & 0.001459171 & 0.000105543 & 0.004395813 \\
		44    & 1.79263E-05 & 1.90663E-05 & 1.90663E-05 & 1.62681E-07 & 5.62215E-05 \\
		64    & 2.34234E-07 & 2.49129E-07 & 2.49129E-07 & 2.50753E-10 & 7.32743E-07 \\
		\bottomrule
	\end{tabular}
\end{table}

\subsubsection{Blockchain Versus FlexiChain}

Observing equation \ref{eq:BC_eq2} and equation \ref{eq:FC_eq8} for the same attack category, it is clear that FlexiChain has more resistance compared to blockchain due to the extra factors integrated to it with ensuring the lightweight operations whereas in blockchain power and time consuming algorithms used. 
 
Observing equation \ref{eq:BC_eq3} and equation \ref{eq:FC_eq9} for the same attack category, it is clear that FlexiChain has more resistance compared to blockchain due to the extra factors integrated to it. The factors guarantee the integrity and security of nodes through the trusted hardware used were if the constructed credentials stolen the trusted hardware credential is secured and separated. The node might need to re-enroll again but the network will be secure. 

Observing equation \ref{eq:BC_eq4} and equation \ref{eq:FC_eq10} for the same attack category, it is clear that FlexiChain has more resistance compared to blockchain due to the extra factors integrated to it. \%51 attack relies mainly in the number of resources nodes or miners, minters, and validators number. Controlling $n/2$ nodes will result in controlling the network using Proof of Work (PoW). In FlexiChain, the case is different, the number of targeted nodes are always equal to the total number of nodes since the authorities are equally distributed. In addition to the extra factors such as trusted hardware credentials and NodeChain with the distributed offline vault.

Observing equation \ref{eq:BC_eq5} and equation \ref{eq:FC_eq11} for the same attack category, it is clear that FlexiChain has more resistance compared to blockchain due to the extra factors integrated to it. Blockchain might be a victim for a quantum brute force attack due to its mere reliance on node credentials as an identity. However, the case in FlexiChain is different the identity is there but secured in two layers the first is the public independent ledger NodeChain and the distributed offline file that resides on trusted hardware with a different set of credentials. if a successful attack acquire the constructed IDs will and trusted module credentials, still will not be able to use the vault since it is offline. 

Equations \ref{eq:BC_eq6} and \ref{eq:FC_eq12} represent the overall probability for the four categories of attack. From the analysis above, the number of nodes is an essential factor in both Blockchain technology and FlexiChain technology. Moreover, digital unique identity and its security and integrity are crucial key for FlexiChain technology. 

\begin{figure}[htb]
\centering
\includegraphics[width=0.90\textwidth]{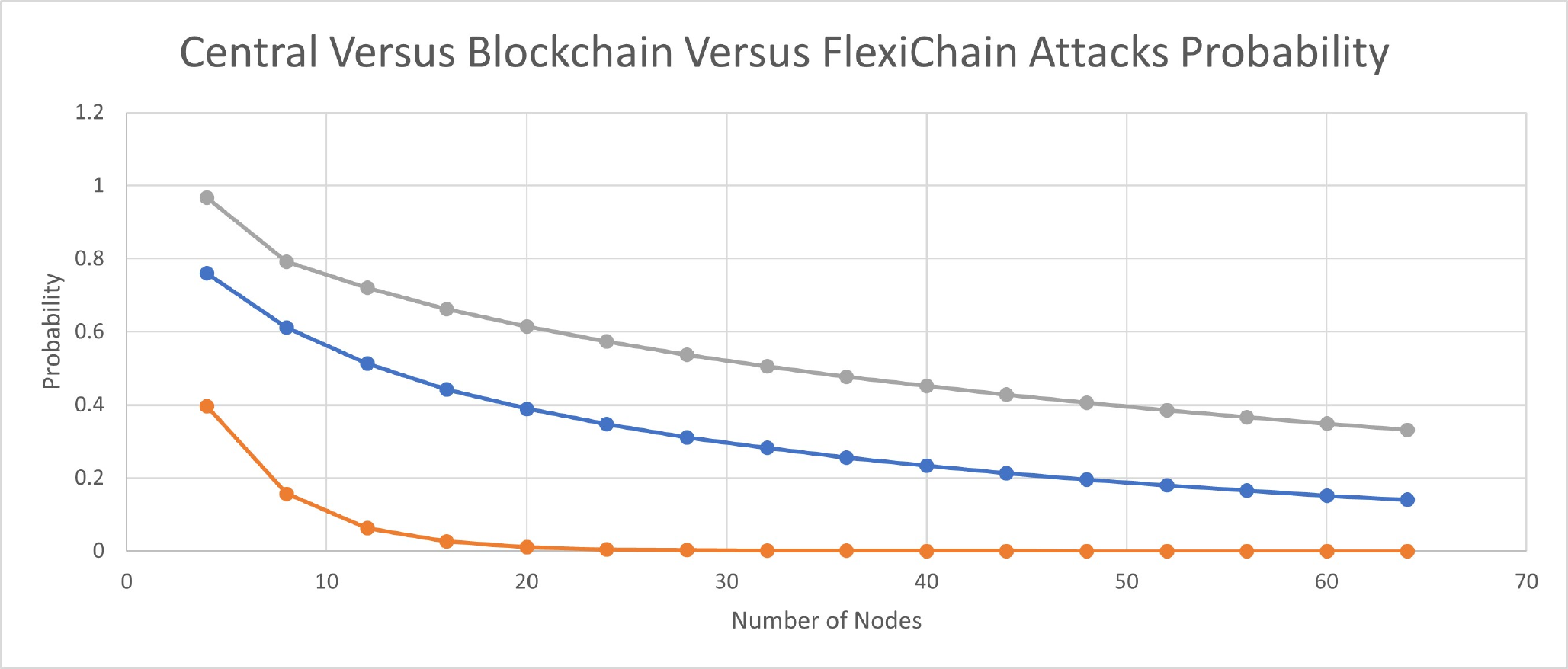}\hspace{1mm}%
\caption{Results Comparison Between Blockchain and FlexiChain}
\label{FIG:ComparisonResultsFlexiChainBlockchainSecurity}
\end{figure}

For equations \ref{eq:BC_eq6} and \ref{eq:FC_eq12}, a simulation has been performed to analyze them using a set of values for each attack based on the possibility of that attack happens and the values range are (0.9-1). In figure  \ref{FIG:ComparisonResultsFlexiChainBlockchainSecurity}, the gray line depict the security analysis for central authority.  The graph shows that the security level for the analyzed categories will be better for a larger number of nodes. However, the increased numbers of participants might raise different categories of attacks not included in this analysis such as transmission attack and DDOS attack. Blockchain security analysis has been presented by the blue graph that illustrates a better security level compared to the centralized paradigm. However, in early stages the security is weaker since the technology depends on the amount of blocks chained and number of nodes. In the case of FlexiChain, the results present a better security level since the technology relies on other methods to compensate the processing difficulty that will be unsuitable for IoT, and CPS applications which speed ip the whole operations. Moreover, the extra factors used in FlexiChain such as UIDS and NodeChain has minimized the probability of all attack categories. Overall, our results illustrate efficiency in term of security based on the above described analysis. Table \ref{TBL:Comparative_Perspective} compares our work with existing frameworks.     
 
\begin{table}[htb]
	\centering
	\caption{Comparative Analysis Between Central Versus Blockchain Versus FlexiChain: This table shows the probabilities acquired from our security analysis for four categories attacks for three scenarios.}
	\label{tab:attackss}
	\begin{tabular}{llll}
		\toprule
		Number of Nodes & Central Probabilities  & Blockchain Probabilities  & FlexiChain Probabilities \\
		\midrule
		4     & 0.9675 & 0.759204611 & 0.396814497 \\
		24    & 0.572781765 & 0.34649071 & 0.004395813 \\
		44    & 0.428023172 & 0.213387706 & 5.62215E-05 \\
		64    & 0.332009476 & 0.139462757 & 7.32743E-07 \\
		\bottomrule
	\end{tabular}
\end{table}

\begin{table*}[htbp]
	\caption{A Comparative Perspective of FlexiChain 2.0 with Previous Works}
	\label{TBL:Comparative_Perspective}
	\centering
	\scriptsize
	\begin{tabular}{p{2.5cm} p{1.5cm} p{3cm} p{1cm} p{1.2cm} p{1.8cm} p{1.5cm}}
		\toprule
		\textbf{Consensus Algorithm} & \textbf{Registration (ms)} & \textbf{Security} & \textbf{Ledger} & \textbf{Miners} & \textbf{Validation} & \textbf{Blockchain Type}\\
		\hline
		Proof of Importance (PoI) \cite{PoI2}                          & Manual    & Digital Signatures, Cryptography & Full & Yes & Accounts Importance & Public\\
		\hline
		Proof of Authority (PoA) \cite{PoA1} \cite{PoA2}               & Manual    & Digital Signatures, Cryptography & Full & Yes & PoS & Permissioned\\
		\hline
		Proof of Authentication (PoAh) \cite{PoAh2}                    & Manual    & Digital Signatures, Cryptography & Full & Yes & Cryptographic & Private\\
		\hline
		Proof of PUF-Enabled Authentication (PoP) \cite{PUFchain}     & Manual    & Digital Signatures, Cryptography, PUF Signature & Full & Yes & Predefined PUF keys verification & Private\\
		\hline
		Proof of Block and Trade (PoBT) \cite{PoBT}                    & Manual    & Digital Signatures, Cryptography & Full  & Yes & Smart Contract and BFT  & Private\\
		\hline
		Proof of Rapid Authentication (PoRa) \textbf{FlexiChain 2.0 (Current)}                                  & Manufactured Trusted Hardware & Digital Signatures, Cryptography, NodeChain, Trusted Hardware & Portion/Full & No & UIDs verification & Public\\
		\bottomrule
	\end{tabular}
\end{table*}

\section{Conclusions}
\label{sec:Conclusions}

FlexiChain 2.0 is an upgraded version of our previous work \cite{Alkhodair2022}. FlexiChain 2.0 has proposed the NodeChain in detail. A chain of narrations has been proposed to authenticate both assets and users both via the distributed vault and utilize the tokenized UID on chain instead of having the real Unique Identification in the NodeChain. Proof of Rapid Authentication is the consensus algorithm used in FlexiChain that relies on guessing UIDS by inserting TUID to match layer. Also, a block can be authenticated more than one time which will create a chain of narrations for who has believed what the block included. NodeChain has created a secure layer above cryptography in order to avoid attacks. Due to the importance of hardware security in limited capabilities nodes, in this paper, we concentrate on enhancing the low level security by merging novel methods and keep the latency at its lowest. FlexiChain is proposed to fulfill Cyber Physical System (CPS) applications in smart cities. Applications such as intelligent transportation, smart healthcare, supply chain management, and data training are targets to test the proposed technology.

\section{Future Directions}
\label{sec:FutureDirections}

An immediate extension of our ongoing research is FlexiChain 3.0 \cite{Alkhodair_MDPI_2023_FlexiChain-3-0}. Effectiveness of the proposed blockchains solutions in the applications domain of smart healthcare is an important are \cite{TCE.2020.3011966, TCE.2019.2940472}. 
For FlexiChain technology, Artificial Intelligence (AI) is a very important factor that can enhance the network operation and increase the automation. Meanwhile, AI needs a secure platform that provide a secure operations. Both Distributed Ledger Technology and AI have a future potential that leverages current services. Moreover, smart contracts could be simulated to provide some access management for users. Finally, it will provide a lightweight programmable asset that could operate and be hosted on FlexiChain and could be a platform for Intelligent Transportation, Data Training, and Supply Chain Management.

\bibliographystyle{IEEEtran}
\bibliography{Bibliography_FlexiChain-2-0}   

\begin{thebibliography}{10}
\providecommand{\url}[1]{#1}
\csname url@samestyle\endcsname
\providecommand{\newblock}{\relax}
\providecommand{\bibinfo}[2]{#2}
\providecommand{\BIBentrySTDinterwordspacing}{\spaceskip=0pt\relax}
\providecommand{\BIBentryALTinterwordstretchfactor}{4}
\providecommand{\BIBentryALTinterwordspacing}{\spaceskip=\fontdimen2\font plus
\BIBentryALTinterwordstretchfactor\fontdimen3\font minus
  \fontdimen4\font\relax}
\providecommand{\BIBforeignlanguage}[2]{{%
\expandafter\ifx\csname l@#1\endcsname\relax
\typeout{** WARNING: IEEEtran.bst: No hyphenation pattern has been}%
\typeout{** loaded for the language `#1'. Using the pattern for}%
\typeout{** the default language instead.}%
\else
\language=\csname l@#1\endcsname
\fi
#2}}
\providecommand{\BIBdecl}{\relax}
\BIBdecl

\bibitem{PoW_Bitcoin}
S.~Nakamoto, ``{Bitcoin: A Peer-to-Peer Electronic Cash System},'' Cryptography
  Mailing list, 2009.

\bibitem{Everything_Blockchain}
D.{Puthal}, N.{Malik}, S.P.{Mohanty}, E.{Kougianos}, and G.{Das}, ``{Everything
  You Wanted to Know About the Blockchain: Its Promise, Components, Processes,
  and Problems},'' \emph{IEEE Consumer Electronics Magazine}, vol.~7, no.~4,
  pp. 6--14, July 2018.

\bibitem{Alkhodair2021}
A.~J. Alkhodair, S.~P. Mohanty, and E.~Kougianos, ``Asid: Accessible secure
  unique identification file based device security in next generation
  blockchains,'' in \emph{2021 IEEE International Conference on Blockchain and
  Cryptocurrency (ICBC)}, 2021, pp. 1--2.

\bibitem{Alkhodair2020}
A.~{Alkhodair}, S.~{Mohanty}, E.~{Kougianos}, and D.~{Puthal}, ``{McPoRA}: A
  multi-chain proof of rapid authentication for post-blockchain based security
  in large scale complex cyber-physical systems,'' in \emph{2020 IEEE Computer
  Society Annual Symposium on VLSI (ISVLSI)}, 2020, pp. 446--451.

\bibitem{Blockchain_framework}
D.~{Puthal}, N.~{Malik}, S.~P. {Mohanty}, E.~{Kougianos}, and C.~{Yang}, ``The
  blockchain as a decentralized security framework [future directions],''
  \emph{IEEE Consumer Electronics Magazine}, vol.~7, no.~2, pp. 18--21, March
  2018.

\bibitem{POS}
S.~King and S.~Nadal, ``{PPCoin: Peer-to-Peer Crypto-Currency with
  Proof-of-Stake},'' https://decred.org/research/king2012.pdf, 2012.

\bibitem{IOTAIOT}
N.~{Živi}, E.~{Kadušić}, and K.~{Kadušić}, ``{Directed Acyclic Graph as
  Tangle: an IoT Alternative to Blockchains},'' in \emph{Proc. 27th
  Telecommunications Forum (TELFOR)}, 2019, pp. 1--3.

\bibitem{DAG1}
D.~{Jungnickel}, \emph{Graphs networks and algorithms}, 4th~ed.\hskip 1em plus
  0.5em minus 0.4em\relax Springer, 2012, no. pp. 92–93.

\bibitem{DAG2}
S.~S. {Skiena}, \emph{The algorithm design manual}, 2nd~ed.\hskip 1em plus
  0.5em minus 0.4em\relax Springer, 2011, no. pp. 495–497.

\bibitem{DAG3}
T.~H. {Cormen}, C.~E. {Leiserson}, R.~L. {Rivest}, and C.~{Stein},
  \emph{{Introduction to Algorithms}}, 2nd~ed.\hskip 1em plus 0.5em minus
  0.4em\relax MIT Press and McGraw-Hill, 2001, no. pp. 552–557.

\bibitem{DAG4}
K.~{Thulasiraman} and M.~N.~S. {Swamy}, \emph{Graphs: Theory and
  Algorithms}.\hskip 1em plus 0.5em minus 0.4em\relax John Wiley and Son, 1992,
  no. p. 118.

\bibitem{HashGraph}
L.~Baird, ``{The Swirlds Hashgraph Consensus Algorithm: Fair, Fast, Byzantine
  Fault Tolerance},'' Swirlds, May 2016.

\bibitem{Hedera}
\BIBentryALTinterwordspacing
L.~B.~M. Harmon and P.~Madsen, ``Hedera: A public hashgraph network and
  governing council,'' p.~97, Aug. 2020. [Online]. Available:
  \url{https://hedera.com/hh_whitepaper_v2.1-20200815.pdf}
\BIBentrySTDinterwordspacing

\bibitem{LeMahieu2015}
\BIBentryALTinterwordspacing
C.~LeMahieu, ``Nano: A feeless distributed cryptocurrency network,'' White
  Paper, 2015. [Online]. Available:
  \url{https://content.nano.org/whitepaper/Nano_Whitepaper_en.pdf}
\BIBentrySTDinterwordspacing

\bibitem{IoTex}
\BIBentryALTinterwordspacing
T.~I. Team, ``Iotexa decentralized network for internet of thingspowered by a
  privacy-centric blockchain,'' White Paper, Jul. 2018. [Online]. Available:
  \url{https://iotex.io/research}
\BIBentrySTDinterwordspacing

\bibitem{VRF}
\BIBentryALTinterwordspacing
S.~M.~M. Rabin and S.~Vadhan, ``Verifiable random functions,'' Tech. Rep.,
  1999. [Online]. Available: \url{https://www.cs.bu.edu/~goldbe/projects/vrf}
\BIBentrySTDinterwordspacing

\bibitem{Nodle}
\BIBentryALTinterwordspacing
Nodle, ``The nodle network:a new economic model tofree the mobile internet,''
  Tech. Rep., Nov. 2021. [Online]. Available:
  \url{https://www.nodle.com/whitepaper-4.2.pdf}
\BIBentrySTDinterwordspacing

\bibitem{IOST}
\BIBentryALTinterwordspacing
T.~I. of~Services~Foundation, ``Internet of services: The next-generation,
  secure, highly scalable ecosystem for online services,'' White Paper, Tech.
  Rep., Dec. 2017. [Online]. Available:
  \url{https://whitepaper.io/document/28/iostoken-whitepaper}
\BIBentrySTDinterwordspacing

\bibitem{PoActivityAndNovelConsensusBased}
Z.~Boreiri and A.~N. Azad, ``A novel consensus protocol in blockchain network
  based on proof of activity protocol and game theory,'' in \emph{2022 8th
  International Conference on Web Research (ICWR)}, 2022, pp. 82--87.

\bibitem{Tangle}
S.~Popov, ``{The Tangle},'' {Jinn Labs}, 2016, version 0.6.

\bibitem{Alkhodair2022}
A.~J. {Alkhodair}, S.~P. {Mohanty}, and E.~{Kougianos}, ``Flexichain: A
  minerless scalable next generation blockchain for rapid data and device
  security in large scale complex cyber-physical systems,'' \emph{Springer
  Nature Computer Science (SN-CS)IEEE Access}, vol.~XX, pp. NN--NN, 2022.

\bibitem{SEP}
\BIBentryALTinterwordspacing
``Apple platform security,'' Apple, Tech. Rep., May 2022. [Online]. Available:
  \url{https://support.apple.com/guide/security/welcome/web}
\BIBentrySTDinterwordspacing

\bibitem{SecurityServiceEnginestoAccelerateEnclavePerformanceinSecureMulticoreProcessors}
J.~Nye and O.~Khan, ``Sse: Security service engines to accelerate enclave
  performance in secure multicore processors,'' \emph{IEEE Computer
  Architecture Letters}, vol.~21, no.~2, pp. 129--132, 2022.

\bibitem{Blockchain_Applications}
W.~{Viriyasitavat}, L.~D. {Xu}, Z.~{Bi}, and D.~{Hoonsopon}, ``Blockchain
  technology for applications in internet of things—mapping from system
  design perspective,'' \emph{IEEE Internet of Things Journal}, vol.~6, no.~5,
  pp. 8155--8168, Oct 2019.

\bibitem{Blockchain_IoT_Edge}
M.~A. {Rahman}, M.~M. {Rashid}, M.~S. {Hossain}, E.~{Hassanain}, M.~F.
  {Alhamid}, and M.~{Guizani}, ``Blockchain and iot-based cognitive edge
  framework for sharing economy services in a smart city,'' \emph{IEEE Access},
  vol.~7, pp. 18\,611--18\,621, 2019.

\bibitem{PUFchain}
S.~P. {Mohanty}, V.~P. {Yanambaka}, E.~{Kougianos}, and D.~{Puthal},
  ``{PUFchain: A Hardware-Assisted Blockchain for Sustainable Simultaneous
  Device and Data Security in the Internet of Everything (IoE)},'' \emph{IEEE
  Consumer Electronics Magazine}, vol.~9, no.~2, pp. 8--16, March 2020.

\bibitem{PuFChain2}
V.~K. V.~V. Bathalapalli, S.~Mohanty, E.~Kougianos, B.~Baniya, and B.~Rout,
  ``Pufchain 2.0: Hardware-assisted robust blockchain for sustainable
  simultaneous device and data security in smart healthcare,'' \emph{SN
  Computer Science}, vol.~3, 06 2022.

\bibitem{Authenticity}
U.~{Guin}, P.~{Cui}, and A.~{Skjellum}, ``{Ensuring Proof-of-Authenticity of
  IoT Edge Devices Using Blockchain Technology},'' in \emph{Proc. IEEE
  International Conference on Internet of Things (iThings) and IEEE Green
  Computing and Communications (GreenCom) and IEEE Cyber, Physical and Social
  Computing (CPSCom) and IEEE Smart Data (SmartData)}, July 2018, pp.
  1042--1049.

\bibitem{PoAh}
D.~{Puthal} and S.~P. {Mohanty}, ``{Proof of Authentication: IoT-Friendly
  Blockchains},'' \emph{IEEE Potentials}, vol.~38, no.~1, pp. 26--29, Jan 2019.

\bibitem{PoAh2}
D.~{Puthal}, S.~P. {Mohanty}, P.~{Nanda}, E.~{Kougianos}, and G.~{Das},
  ``{Proof-of-Authentication for Scalable Blockchain in Resource-Constrained
  Distributed Systems},'' in \emph{Proc. IEEE International Conference on
  Consumer Electronics (ICCE)}, 2019, pp. 1--5.

\bibitem{PoBT}
S.~{Biswas}, K.~{Sharif}, F.~{Li}, S.~{Maharjan}, S.~P. {Mohanty}, and
  Y.~{Wang}, ``{PoBT: A Lightweight Consensus Algorithm for Scalable IoT
  Business Blockchain},'' \emph{IEEE Internet of Things Journal}, vol.~7,
  no.~3, pp. 2343--2355, March 2020.

\bibitem{POTrust}
J.~{Zou}, B.~{Ye}, L.~{Qu}, Y.~{Wang}, M.~A. {Orgun}, and L.~{Li}, ``A
  proof-of-trust consensus protocol for enhancing accountability in
  crowdsourcing services,'' \emph{IEEE Transactions on Services Computing},
  vol.~12, no.~3, pp. 429--445, May 2019.

\bibitem{Fortified}
B.~S. Egala, A.~K. Pradhan, V.~Badarla, and S.~P. Mohanty, ``Fortified-chain: A
  blockchain-based framework for security and privacy-assured internet of
  medical things with effective access control,'' \emph{IEEE Internet of Things
  Journal}, vol.~8, no.~14, pp. 11\,717--11\,731, 2021.

\bibitem{GlobalChain}
S.~Biswas, K.~Sharif, F.~Li, A.~K. Bairagi, Z.~Latif, and S.~P. Mohanty,
  ``Globechain: An interoperable blockchain for global sharing of healthcare
  data—a covid-19 perspective,'' \emph{IEEE Consumer Electronics Magazine},
  vol.~10, no.~5, pp. 64--69, 2021.

\bibitem{DAAC}
S.~Biswas, K.~Sharif, F.~Li, I.~Alam, and S.~Mohanty, ``Daac: Digital asset
  access control in a unified blockchain based e-health system,'' \emph{IEEE
  Transactions on Big Data}, pp. 1--1, 2020.

\bibitem{EPOW_HealthRecords}
G.~Gunanidhi and R.~Krishnaveni, ``Improved security blockchain for iot based
  healthcare monitoring system,'' in \emph{2022 Second International Conference
  on Artificial Intelligence and Smart Energy (ICAIS)}, 2022, pp. 1244--1247.

\bibitem{ABlockchain_empoweredFederatedLearninginHealthcare_basedCyberPhysicalSystems}
Y.~Liu, W.~Yu, Z.~Ai, G.~Xu, L.~Zhao, and Z.~Tian, ``A blockchain-empowered
  federated learning in healthcare-based cyber physical systems,'' \emph{IEEE
  Transactions on Network Science and Engineering}, pp. 1--1, 2022.

\bibitem{smartHealthcareDiseaseManagment}
K.~Azbeg, O.~Ouchetto, and S.~J. Andaloussi, ``Access control and
  privacy-preserving blockchain-based system for diseases management,''
  \emph{IEEE Transactions on Computational Social Systems}, pp. 1--13, 2022.

\bibitem{PharmaChain}
A.~Bapatla, S.~Mohanty, and E.~Kougianos, ``Pharmachain: A blockchain to ensure
  counterfeit free pharmaceutical supply chain,'' 02 2022.

\bibitem{CryptoPharmacy}
G.~Subramanian, A.~Sreekantanthampy, N.~Valbosco, and B.~Ramnani, ``Crypto
  pharmacy – digital medicine: A mobile application integrated with hybrid
  blockchain to tackle the issues in pharma supply chain,'' \emph{IEEE Open
  Journal of the Computer Society}, vol.~PP, pp. 1--1, 01 2021.

\bibitem{GaRuDa}
R.~Gupta, P.~Bhattacharya, S.~Tanwar, N.~Kumar, and S.~Zeadally, ``Garuda: A
  blockchain-based delivery scheme using drones for healthcare 5.0
  applications,'' \emph{IEEE Internet of Things Magazine}, vol.~4, no.~4, pp.
  60--66, 2021.

\bibitem{Blockchainseverywhereausecaseofblockchaininthepharmasupplychain}
T.~Bocek, B.~B. Rodrigues, T.~Strasser, and B.~Stiller, ``Blockchains
  everywhere - a use-case of blockchains in the pharma supply-chain,'' in
  \emph{2017 IFIP/IEEE Symposium on Integrated Network and Service Management
  (IM)}, 2017, pp. 772--777.

\bibitem{TraceabilityofcounterfeitmedicinesupplychainthroughBlockchain}
R.~Kumar and R.~Tripathi, ``Traceability of counterfeit medicine supply chain
  through blockchain,'' in \emph{Proc. 11th International Conference on
  Communication Systems and Networks (COMSNETS)}, 2019, pp. 568--570.

\bibitem{Drugledger}
Y.~Huang, J.~Wu, and C.~Long, ``Drugledger: A practical blockchain system for
  drug traceability and regulation,'' in \emph{Proc. IEEE International
  Conference on Internet of Things (iThings) and IEEE Green Computing and
  Communications (GreenCom) and IEEE Cyber, Physical and Social Computing
  (CPSCom) and IEEE Smart Data (SmartData)}, 2018, pp. 1137--1144.

\bibitem{Blockchain_Based_DeepLearning_CPS}
S.~Rathore and J.~H. Park, ``A blockchain-based deep learning approach for
  cyber security in next generation industrial cyber-physical systems,''
  \emph{IEEE Transactions on Industrial Informatics}, vol.~17, no.~8, pp.
  5522--5532, 2021.

\bibitem{DeepReinforcement_Edge_Blockchain}
Y.~He, Y.~Wang, C.~Qiu, Q.~Lin, J.~Li, and Z.~Ming, ``Blockchain-based edge
  computing resource allocation in iot: A deep reinforcement learning
  approach,'' \emph{IEEE Internet of Things Journal}, vol.~8, no.~4, pp.
  2226--2237, 2021.

\bibitem{TactileInternetforAutonomousVehiclesLatencyandReliabilityAnalysis}
S.~Tanwar, S.~Tyagi, I.~Budhiraja, and N.~Kumar, ``Tactile internet for
  autonomous vehicles: Latency and reliability analysis,'' \emph{IEEE Wireless
  Communications}, vol.~26, no.~4, pp. 66--72, 2019.

\bibitem{ArtificialIntelligenceEmpoweredEdgeComputingandCachingforInternetofVehicles}
Y.~Dai, D.~Xu, S.~Maharjan, G.~Qiao, and Y.~Zhang, ``Artificial intelligence
  empowered edge computing and caching for internet of vehicles,'' \emph{IEEE
  Wireless Communications}, vol.~26, no.~3, pp. 12--18, 2019.

\bibitem{JointPricingandPowerAllocationforMultibeamSatelliteSystemsWithDynamicGameModel}
F.~Li, K.-Y. Lam, X.~Liu, J.~Wang, K.~Zhao, and L.~Wang, ``Joint pricing and
  power allocation for multibeam satellite systems with dynamic game model,''
  \emph{IEEE Transactions on Vehicular Technology}, vol.~67, no.~3, pp.
  2398--2408, 2018.

\bibitem{SpectralEfficiencyEnhancementinSatelliteMobileCommunicationsAGameTheoreticalApproach}
F.~Li, K.-Y. Lam, H.-H. Chen, and N.~Zhao, ``Spectral efficiency enhancement in
  satellite mobile communications: A game-theoretical approach,'' \emph{IEEE
  Wireless Communications}, vol.~27, no.~1, pp. 200--205, 2020.

\bibitem{BlockchainfortheInternetofVehiclesTowardsIntelligentTransportationSystemsASurvey}
M.~B. Mollah, J.~Zhao, D.~Niyato, Y.~Guan, S.~Sun, K.-Y. Lam, and L.~Koh,
  ``Blockchain for the internet of vehicles towards intelligent transportation
  systems: A survey,'' \emph{IEEE Internet of Things Journal}, vol.~PP, pp.
  1--28, 10 2020.

\bibitem{BlockchainforSecureandEfficientDataSharinginVehicularEdgeComputingandNetworks}
J.~Kang, R.~Yu, X.~Huang, M.~Wu, S.~Maharjan, S.~Xie, and Y.~Zhang,
  ``Blockchain for secure and efficient data sharing in vehicular edge
  computing and networks,'' \emph{IEEE Internet of Things Journal}, vol.~6,
  no.~3, pp. 4660--4670, 2019.

\bibitem{DrivMan}
U.~Javaid, M.~N. Aman, and B.~Sikdar, ``Drivman: Driving trust management and
  data sharing in vanets with blockchain and smart contracts,'' in \emph{Proc.
  IEEE 89th Vehicular Technology Conference (VTC2019-Spring)}, 2019, pp. 1--5.

\bibitem{TraceableandAuthenticatedKeyNegotiationsviaBlockchainforVehicularCommunications}
Y.~Chen, X.~Hao, W.~Ren, and Y.~Ren, ``Traceable and authenticated key
  negotiations via blockchain for vehicular communications,'' \emph{Mobile
  Information Systems}, vol. 2019, pp. 1--10, 12 2019.

\bibitem{ASecureandEfficientBlockchainBasedDataTradingApproachforInternetofVehicles}
C.~Chen, J.~Wu, H.~Lin, W.~Chen, and Z.~Zheng, ``A secure and efficient
  blockchain-based data trading approach for internet of vehicles,'' \emph{IEEE
  Transactions on Vehicular Technology}, vol.~68, no.~9, pp. 9110--9121, 2019.

\bibitem{ComputingResourceTradingforEdgeCloudAssistedInternetofThings}
Z.~Li, Z.~Yang, and S.~Xie, ``Computing resource trading for
  edge-cloud-assisted internet of things,'' \emph{IEEE Transactions on
  Industrial Informatics}, vol.~PP, pp. 1--1, 02 2019.

\bibitem{BlockchainEmpoweredResourceTradinginMobileEdgeComputingandNetworks}
G.~Qiao, S.~Leng, H.~Chai, A.~Asadi, and Y.~Zhang, ``Blockchain empowered
  resource trading in mobile edge computing and networks,'' in \emph{ICC 2019 -
  2019 IEEE International Conference on Communications (ICC)}, 2019, pp. 1--6.

\bibitem{PrivacyPreservingSmartParkingSystemUsingBlockchainandPrivateInformationRetrieval}
W.~A. Amiri, M.~Baza, K.~Banawan, M.~Mahmoud, W.~Alasmary, and K.~Akkaya,
  ``Privacy-preserving smart parking system using blockchain and private
  information retrieval,'' in \emph{2019 International Conference on Smart
  Applications, Communications and Networking (SmartNets)}, 2019, pp. 1--6.

\bibitem{WhenDoWeNeedAblockchain}
D.~Puthal, S.~P. Mohanty, E.~Kougianos, and G.~Das, ``When do we need the
  blockchain?'' \emph{IEEE Consumer Electronics Magazine}, vol.~10, no.~2, pp.
  53--56, 2021.

\bibitem{Dedeoglu2020}
V.~Dedeoglu, A.~Dorri, R.~Jurdak, R.~A. Michelin, R.~C. Lunardi, S.~S. Kanhere,
  and A.~F. Zorzo, ``A journey in applying blockchain for cyberphysical
  systems,'' in \emph{2020 International Conference on COMmunication Systems
  NETworkS (COMSNETS)}, 2020, pp. 383--390.

\bibitem{IoT_meets_Blockchain}
B.~{Cao}, Y.~{Li}, L.~{Zhang}, L.~{Zhang}, S.~{Mumtaz}, Z.~{Zhou}, and
  M.~{Peng}, ``When internet of things meets blockchain: Challenges in
  distributed consensus,'' \emph{IEEE Network}, pp. 1--7, 2019.

\bibitem{Safeguard}
N.~{Kolokotronis}, K.~{Limniotis}, S.~{Shiaeles}, and R.~{Griffiths},
  ``{Secured by Blockchain: Safeguarding Internet of Things Devices},''
  \emph{IEEE Consumer Electronics Magazine}, vol.~8, no.~3, pp. 28--34, May
  2019.

\bibitem{SCRYPT}
\BIBentryALTinterwordspacing
C.~Percival, ``Stronger key derivation via sequential memory-hard functions,''
  Tarsnap, 2012. [Online]. Available:
  \url{https://www.tarsnap.com/scrypt/scrypt.pdf}
\BIBentrySTDinterwordspacing

\bibitem{PoI2}
K.~Au, ``{Tracing Back Stolen Cryptocurrency (XEM) From Japan's Coincheck},''
  Forbes.

\bibitem{PoA1}
\emph{Parity: Fast, light, robust Ethereum implementation, Parity Technologies,
  2017-12-12, retrieved 2017-12-12}.

\bibitem{PoA2}
\emph{Gavin, Wood (November 2015). "PoA Private Chains". Github.},
  https://github.com/poanetwork/wiki/wiki/POA-Network-Whitepaper.

\bibitem{Alkhodair_MDPI_2023_FlexiChain-3-0}
A.~J. Alkhodair, S.~P. Mohanty, and E.~Kougianos, ``Flexichain 3.0: Distributed
  ledger technology based intelligent transportation for vehicular digital
  asset exchange in smart cities,'' \emph{MDPI Sensors}, p. Accepted on 17 Apr
  2023, 2023.

\bibitem{TCE.2020.3011966}
\BIBentryALTinterwordspacing
A.~M. Joshi, P.~Jain, S.~P. Mohanty, and N.~Agrawal, ``{iGLU 2.0: A New
  Wearable for Accurate Non-Invasive Continuous Serum Glucose Measurement in
  IoMT Framework},'' \emph{{IEEE} Trans. Consumer Electron.}, vol.~66, no.~4,
  pp. 327--335, 2020. [Online]. Available:
  \url{https://doi.org/10.1109/TCE.2020.3011966}
\BIBentrySTDinterwordspacing

\bibitem{TCE.2019.2940472}
\BIBentryALTinterwordspacing
L.~Rachakonda, S.~P. Mohanty, E.~Kougianos, and P.~Sundaravadivel,
  ``{Stress-Lysis: A DNN-Integrated Edge Device for Stress Level Detection in
  the IoMT},'' \emph{{IEEE} Trans. Consumer Electron.}, vol.~65, no.~4, pp.
  474--483, 2019. [Online]. Available:
  \url{https://doi.org/10.1109/TCE.2019.2940472}
\BIBentrySTDinterwordspacing

\end{thebibliography}

\begin{IEEEbiography}
[{\includegraphics[height=1.25in, keepaspectratio]{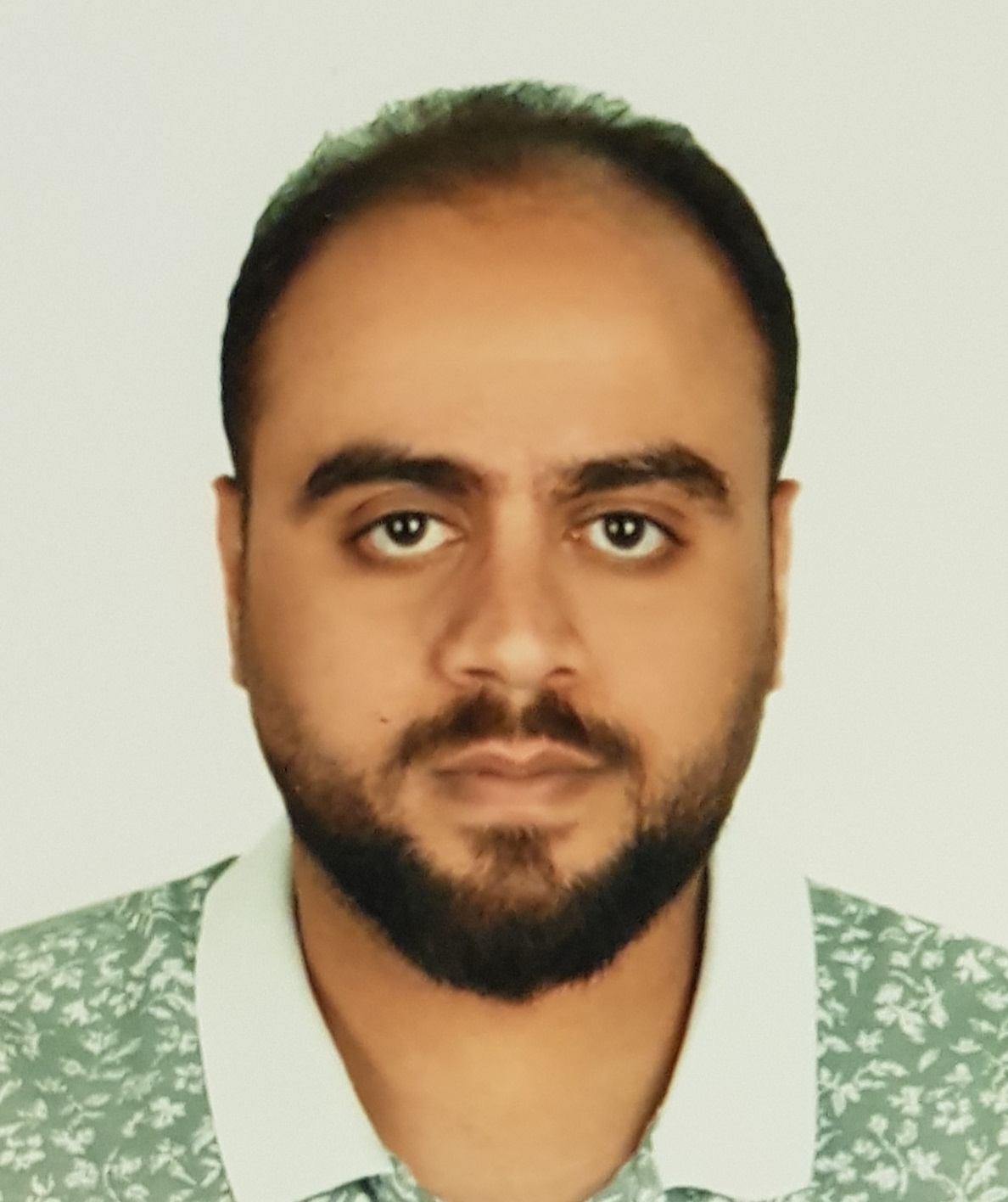}}]
{Ahmad J. Alkhodair} received a bachelor's degree (Honors) in computer engineering from Fahad bin Sultan University (FBSU), Saudi Arabia-Tabuk, in 2012 and master's in computer engineering in 2017 from Denver University (DU), Colorado-Denver, USA. Currently a Ph.D. candidate in the research group at Smart Electronics Systems Laboratory (SESL) at Computer Science and Engineering at the University of North Texas, Denton, TX. Research interests include Distributed Ledger Technology (DLT), Cyber Physical Systems (CPS), Artificial Intelligence (AI) in Smart Cities. 
\end{IEEEbiography}

\begin{IEEEbiography}
[{\includegraphics[height=1.25in, keepaspectratio]{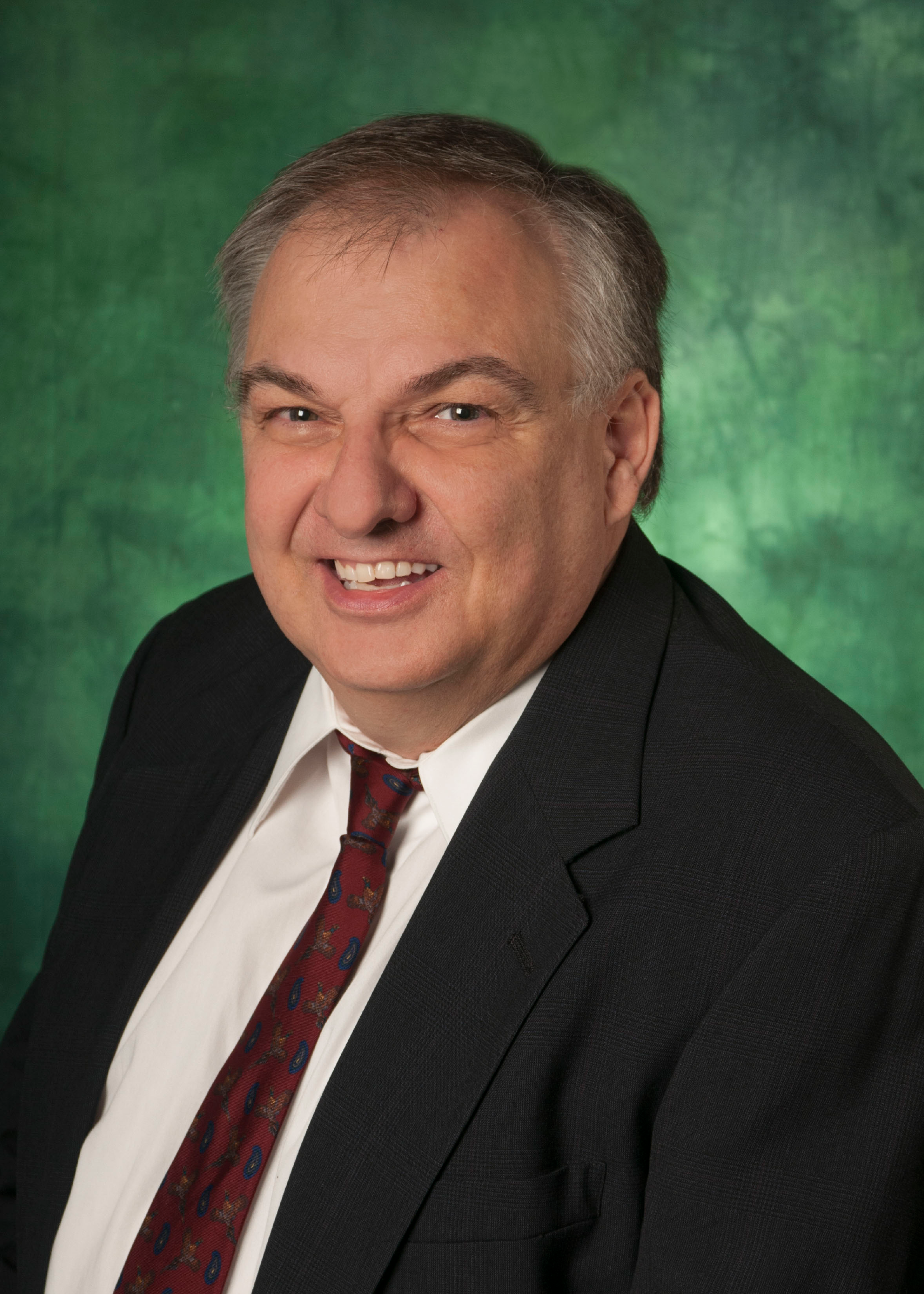}}]
{Elias Kougianos} received a BSEE from the University of Patras, Greece in 1985 and an MSEE in 1987, an MS in Physics in 1988 and a Ph.D. in EE in 1997, all from Louisiana State University. From 1988 through 1998 he was with Texas Instruments, Inc., in Houston and Dallas, TX. In 1998 he joined Avant! Corp. (now Synopsys) in Phoenix, AZ as a Senior Applications engineer and in 2000 he joined Cadence Design Systems, Inc., in Dallas, TX as a Senior Architect in Analog/Mixed-Signal Custom IC design. He has been at UNT since 2004. He is a Professor in the Department of Electrical Engineering, at the University of North Texas (UNT), Denton, TX. His research interests are in the area of Analog/Mixed-Signal/RF IC design and simulation and in the development of VLSI architectures for multimedia applications. He is an author of over 200 peer-reviewed journal and conference publications.
\end{IEEEbiography}

\begin{IEEEbiography}
[{\includegraphics[height=1.25in, keepaspectratio]{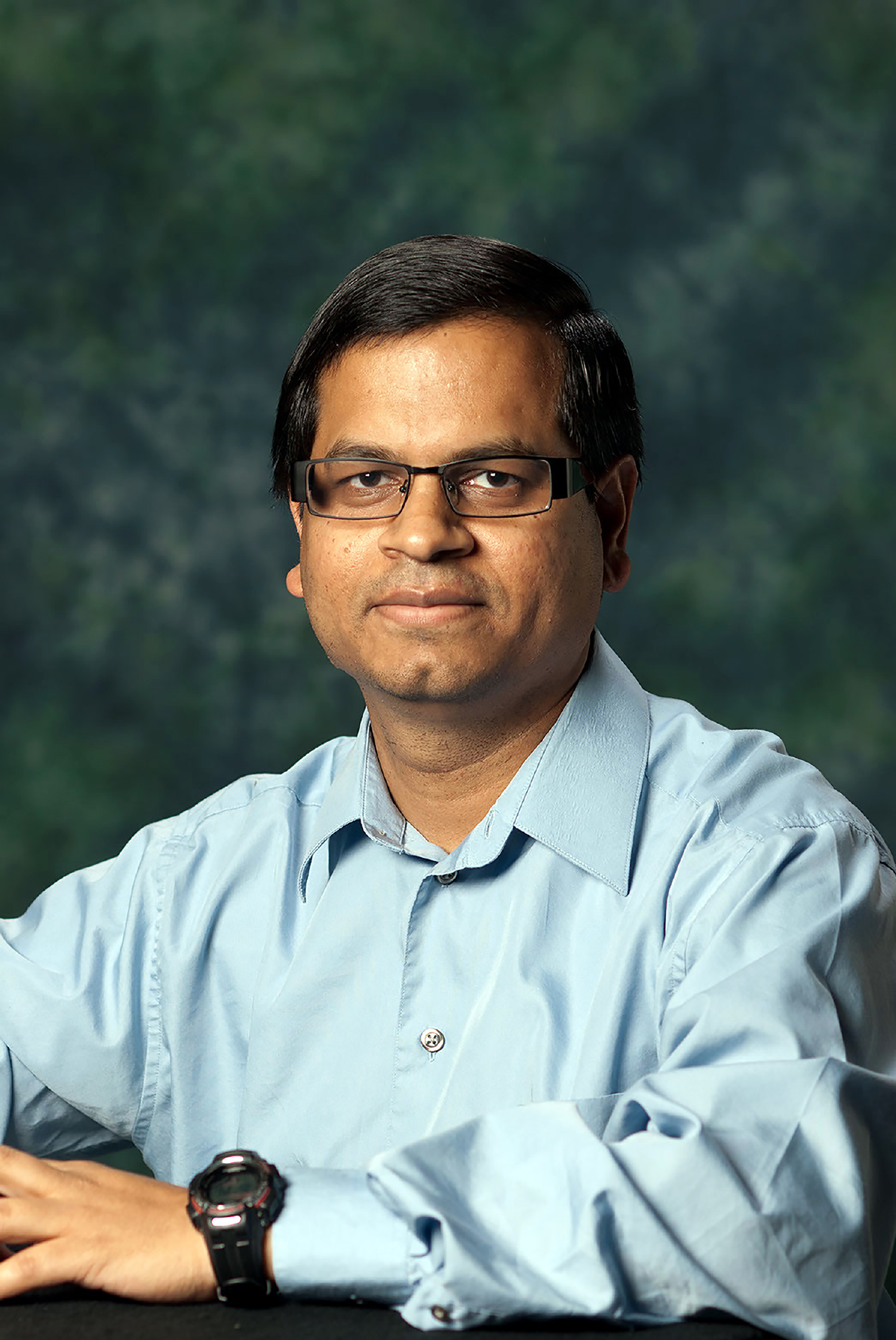}}]
{Saraju P. Mohanty} received the bachelor's degree (Honors) in electrical engineering from the Orissa University of Agriculture and Technology, Bhubaneswar, in 1995, the master's degree in Systems Science and Automation from the Indian Institute of Science, Bengaluru, in 1999, and the Ph.D. degree in Computer Science and Engineering from the University of South Florida, Tampa, in 2003. He is a Professor with the University of North Texas. His research is in ``Smart Electronic Systems'' which has been funded by National Science Foundations (NSF), Semiconductor Research Corporation (SRC), U.S. Air Force, IUSSTF, and Mission Innovation. He has authored 450 research articles, 5 books, and 9 granted and pending patents. His Google Scholar h-index is 49 and i10-index is 217 with 11,000 citations. He is regarded as a visionary researcher on Smart Cities technology in which his research deals with security and energy aware, and AI/ML-integrated smart components. He introduced the Secure Digital Camera (SDC) in 2004 with built-in security features designed using Hardware Assisted Security (HAS) or Security by Design (SbD) principle. He is widely credited as the designer for the first digital watermarking chip in 2004 and first the low-power digital watermarking chip in 2006. He is a recipient of 16 best paper awards, Fulbright Specialist Award in 2020, IEEE Consumer Electronics Society Outstanding Service Award in 2020, the IEEE-CS-TCVLSI Distinguished Leadership Award in 2018, and the PROSE Award for Best Textbook in Physical Sciences and Mathematics category in 2016. He has delivered 18 keynotes and served on 14 panels at various International Conferences. He has been serving on the editorial board of several peer-reviewed international transactions/journals, including IEEE Transactions on Big Data (TBD), IEEE Transactions on Computer-Aided Design of Integrated Circuits and Systems (TCAD), IEEE Transactions on Consumer Electronics (TCE), and ACM Journal on Emerging Technologies in Computing Systems (JETC). He has been the Editor-in-Chief (EiC) of the IEEE Consumer Electronics Magazine (MCE) during 2016-2021. He served as the Chair of Technical Committee on Very Large Scale Integration (TCVLSI), IEEE Computer Society (IEEE-CS) during 2014-2018 and on the Board of Governors of the IEEE Consumer Electronics Society during 2019-2021. He serves on the steering, organizing, and program committees of several international conferences. He is the steering committee chair/vice-chair for the IEEE International Symposium on Smart Electronic Systems (IEEE-iSES), the IEEE-CS Symposium on VLSI (ISVLSI), and the OITS International Conference on Information Technology (OCIT). He has mentored 2 post-doctoral researchers, and supervised 14 Ph.D. dissertations, 26 M.S. theses, and 18 undergraduate projects.
\end{IEEEbiography}

\end{document}